\documentclass[amsmath,amssymb,14pt]{article}

\usepackage{graphicx, epsfig}

\usepackage{amssymb,amsmath,color}

\usepackage{dcolumn}
\usepackage{bm}
\newcommand{\be}{\begin{equation}}
\newcommand{\ee}{\end{equation}}
\newcommand{\beqs}{\begin{eqnarray}}
\newcommand{\eeqs}{\end{eqnarray}}

\begin{document}
\title{\textsc{
A Perturbative Approach to the tunneling Phenomena} }

\author{\centerline {Fatih Erman$^1$, O. Teoman Turgut$^{2, 3}$}
\\\and
 {\scriptsize{$^1$
Department of Mathematics, \.{I}zmir Institute of Technology, Urla,
35430, \.{I}zmir, Turkey}}
\\\and
 {\scriptsize{$^2$
Department of Physics, Bo\u{g}azi\c{c}i University, 34342 Bebek, \.{I}stanbul, Turkey}} 
\\\and
 {\scriptsize{$^3$
Department of Physics, Carnegie Mellon University, Pittsburgh, PA 15213, USA}}
\\
{\scriptsize{E-mail: fatih.erman@gmail.com, turgutte@boun.edu.tr}}}

\maketitle

\begin{abstract}

The double-well potential is a good example, where we can compute the splitting in the bound state energy of the system due to the tunneling effect with various methods,  namely WKB or instanton calculations. All these methods are non-perturbative and there is a common belief that it is difficult to find the splitting in the energy due to the barrier penetration from a perturbative analysis. However, we will illustrate by explicit examples containing singular potentials (e.g., Dirac delta potentials supported by points and curves and their relativistic extensions) that it is possible to find the splitting in the bound state energies by developing some kind of perturbation method.

\end{abstract}


\section{Introduction}

Most real quantum mechanical systems can not be solved exactly and we usually apply some approximation methods, the most common one being perturbation theory, to get information about the energy levels and scattering amplitudes. However, not all quantum systems can be analyzed by perturbative methods. There are various problems for which one can not deduce any information by simply using perturbation theory since these problems are inherently non-perturbative phenomena like the formation of bound states or penetration through  a  potential barrier. For such non-perturbative phenomenon,  other tools, such as  WKB \cite{LL, Das} or instanton calculations \cite{Coleman}, are particularly useful. The particle moving in a one-dimensional anharmonic potential $V(x)= {\lambda^2 \over 8} (x^2-a^2)^2$ is a classic example, where we can study the barrier penetration through the WKB analysis.  
\begin{figure}[h!]
\centering
\includegraphics[scale=0.5]{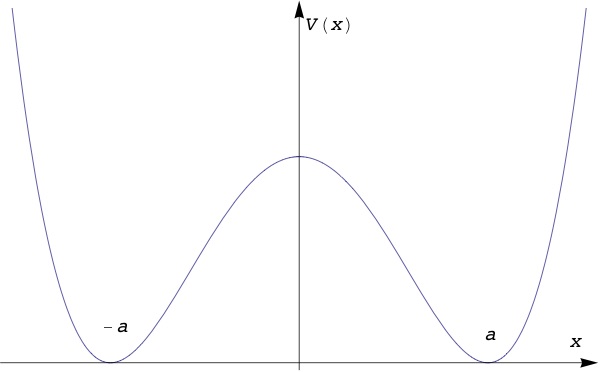}
\caption{Anharmonic Potential} \label{anharmonic potential}
\end{figure}
When the energy scale determined by the length scale $a$ is extremely small compared with the binding energy of the system, i.e., $\hbar^2/2ma^2 << E_B$, or $\lambda a^2>>1$,  the potential separates into two symmetrical wells with a very high barrier (see Fig. \ref{anharmonic potential}). In this extreme regime, as  a first approximation, each well has separately  quantized energy levels and these energy levels are degenerate due to the symmetry. However, once the large but finite value of the coupling constant $\lambda$ is taken into account, the particle initially confined to one well can tunnel to the other well so the degeneracy in the energy levels disappear. The splitting in the resulting energy levels  (between the true ground state and the first excited level due to the tunneling)  is given by \cite{LL, Das}
\begin{equation}
\delta E = E_2-E_1 \approx {4 e \over \pi} \sqrt{m \hbar} \omega^{3/2} a \exp\left(- {1 \over \hbar} S_0 \right)
\end{equation}
where $S_0= {2 m \omega a^2 \over 3}$ and $\omega^2= {\lambda^2 a^2 \over m}$.
The above exponentially decaying factor with respect to the separation between the wells illustrates the tunneling effect. The true ground state corresponds to a symmetric combination and the excited level corresponds to the anti-symmetric combination of the  WKB corrected wave functions.

Among the exactly solvable potentials in quantum mechanics, Dirac delta well potentials are the most well-known text book example \cite{Galindo Pascual1}. Moreover, it has been studied extensively in mathematical physics literature from  different point of views, in particular in the context of self-adjoint extension of symmetric operators \cite{RS2}. Although it is easier to define it rigorously in one dimension through the quadratic forms, one possible way to define it in higher dimensions is to consider the free symmetric Hamiltonian defined on a dense domain excluding the point, where the support of the Dirac delta function is located, and then apply the self-adjoint extension techniques developed by J. Von Neumann (see the monograph \cite{Albeverio} for the details and also for the historical development with extensive literature in the subject). Then, the formal (or heuristic) definition of one-dimensional Dirac delta potentials in the physics literature is understood as the one particular choice among the four parameter family of the self-adjoint operators, where the matching conditions of the wave function are just obtained from the boundary conditions (which define the domain of our self-adjoint operator) constructed through the extension theory. Another way to introduce these point interactions uses  the resolvent method, developed by  M. Krein, and it is based on the observation that for such type of potentials the resolvent can be found explicitly and  expressed via the so-called Krein's formula \cite{Albeverio Kurasov}. Within this approach, the Hamiltonian for point interaction (in two and three dimensions)  is first approximated (regularized) by a properly chosen sequence of self-adjoint operators $H_{\epsilon}$ and then the coupling constant (or strength) of the potential is assumed to be  a function of the parameter $\epsilon$ in such a way that one obtains  a non-trivial limit. This convergence is actually in the strong resolvent sense, so the limit operator is self-adjoint \cite{Berezin Faddeev}. Since the Dirac delta potentials in two and three dimensions require renormalization, it is usually considered as a toy model for the renormalization originally developed in quantum field theories and it helps us to better understand  various ideas in field theory such as renormalization group and asymptotic freedom \cite{Huang, Jackiw, GT, MG}. Furthermore, point like Dirac delta interactions have also been  extended to various more general cases. For our approach, to illustrate the main ideas, we are mainly concerned with the delta potentials supported by points on flat and hyperbolic manifolds \cite{AltunkaynakErmanTurgut, ErmanTurgut, ErmannumberH}, and delta potentials supported  by curves in flat spaces, and its various relativistic extensions in flat spaces \cite{ExnerKondej1, Caglar1, ermangadellauncu, BurakTeo}.

In this paper, we  explicitly demonstrate  for a class of  singular potential problems that the splitting in the energy levels due to the tunneling can be realized by simply developing some kind of perturbation theory. We have two basic assumptions here: 1) Binding energies of individual Dirac delta potentials are all different. Otherwise we need to employ degenerate perturbation theory. Actually, we briefly discuss a particular degenerate case, namely the two center case  to compare with the double well potential.
2) The support of singular interactions  are sufficiently separated from one another, as a result the bound state wave functions decay rapidly over the distances between them.

All the findings about the splitting in the bound state energies for singular potentials on hyperbolic manifolds treated here   
could be applied to the two dimensional systems such as graphite sheets. We can model impurities in these systems as attractive centers in some approximation and these sheets can be put in various shapes. This is especially true for surfaces with variable sectional curvature which  is not completely negative.
The negatively curved surfaces, of course, cannot be realized as embedded surfaces in three dimensions due to Hilbert's well-known theorem. Nevertheless, we may envisage these models  as an effective description of  unusual quasi-particle states of some  two dimensional systems. Due to the interactions, the system may develop a gap in the spectrum and the effective description may well be best understood through a negative sectional curvature space. Models related to point interactions on Lobachevsky plane have been studied from variety of different perspectives in \cite{AlbeverioExnerGeyler, BruningGeyler}. The point interactions can be extended on more general class of manifolds as well \cite{BGP}. In particular, 
they have been studied on some particular surfaces in $\mathbb{R}^3$, namely on the infinite planar strip as a natural model for quantum wires containing impurities \cite{Exner} and on the torus \cite{Rudnick torus}. A more heuristic approach for point interactions on Riemannian manifolds has been constructed through the heat kernel in  \cite{AltunkaynakErmanTurgut, ErmanTurgut}.
The physical motivation behind studying the Dirac delta potentials supported by curves is based on the need for modelling semiconductor wires \cite{Exner2}. They could be considered as  toy models for electrons confined to narrow tube-like regions.

The paper is organized as follows: In Section \ref{Krein's Formulae for Singular Interactions}, we formally summarize the resolvent formulae, called Krein's formulae, for Hamiltonians perturbed by singular potentials including Dirac delta potentials supported by points and curves. The principal matrix for each case is given explicitly. The relativistic and the field theoretical extension of it has been also reviewed in the subsections of this section. In Section \ref{Bound State Spectrum}, we briefly discuss the analytic structure of the principal matrix and the bound state spectrum for such  singular interactions. In Section \ref{Off-Diagonal Terms of the Principal Matrix in the Tunneling Regime}, we discuss how the off-diagonal terms of the principal matrix change in the tunneling regime. 
Section \ref{Splitting in Bound State Energies through Perturbation Theory} and Section \ref{Explicit Examples for the Splitting in the Energy} contain the formulation of the perturbative analysis and explicit calculations of the splitting in the bound state energy when these singular potentials are placed far away from each other, which is the main result of the paper. We finally 
 discuss the degenerate case and wave functions, and compare these results results with the exact result in   
Section \ref{Degenerate Case and Wave Functions for Point Interactions} to get a feeling for the accuracy of our approximation.

\section{Krein's Formulae for Free Hamiltonians Perturbed by Singular Interactions} 
\label{Krein's Formulae for Singular Interactions}

Before we are going to discuss the perturbative analysis of singular interactions \textit{for large separations of the supports of these potentials}, we first present the basic results  about our formulation of the singular Hamiltonians. In this paper, we are mainly concerned with the Dirac delta potentials supported by finitely many points and finitely many curves in flat spaces, and their extension to the hyperbolic manifolds. Moreover, we  also consider some relativistic extensions of these singular interactions.

Since we study the spectral properties of different kinds of Dirac delta potentials, we first introduce the notation for  Dirac delta functions of interest.  The Dirac delta distribution $\delta_a$ supported by a point $\mathbf
{a}$ in $\mathbb{R}^n$ is defined as a continuous linear functional whose action on the test functions $\psi$ is given by 
\begin{equation}
\langle \delta_a, \psi \rangle =\psi(\mathbf{a}) \;. \label{pointdirac}
\end{equation}
Similarly, Dirac delta distribution $\delta_{\gamma}$ supported by a curve $\Gamma$ in $\mathbb{R}^n$ is defined as a continuous linear functional whose action on the test functions $\psi$ is given by \cite{Appel}
\begin{equation}
\langle \delta_{\gamma}, \psi \rangle = \int_{\Gamma} d s \;  \psi(\gamma(s)) \label{curvedelta} \;.
\end{equation}
The left hand sides in the definitions (\ref{pointdirac}) and (\ref{curvedelta}) can be expressed in the Dirac's bra-ket notation, most common in physics literature, as $\langle \mathbf{a} | \psi \rangle$ and $\langle \mathbf{\gamma} | \psi \rangle$, respectively.

As we have already emphasized in the introduction, there are several ways to define rigorously the Hamiltonian for Dirac delta potentials. Here, we start with a finite rank perturbations of self-adjoint free Hamiltonian $H_0$ (e.g., $H_0=P^2/2m$ in the non-relativistic case and $H_0=\sqrt{P^2 +m^2}$ in the semi-relativistic case):
\begin{equation}
H= H_0 - \sum_{i=1}^{N} \lambda_i \langle \varphi_i, . \rangle \; \varphi_i \;,
\label{finite rank perturbed Hamiltonian}
\end{equation}  
where $\varphi_i \in \mathcal{H}$ and $\langle . , . \rangle$ denotes the sesqui-linear inner product in the Hilbert space $\mathcal{H}$. Then, it is well-known that the resolvent of $H$ can be explicitly found in terms of the resolvent of the free part by simply solving the inhomogenous equation \cite{Albeverio Kurasov}
\begin{equation}
(H-z) \psi = \rho \;, \label{inhomogenous eq}
\end{equation}
for a given $\rho \in \mathcal{H}$ and $\psi \in D(H_0)=D(H)$. Here $D$ stands for the domain of the operator and we assume that $\Im(z)>0$. It is well-known that $H$ is self-adjoint on $D(H_0)$ due to the Kato-Rellich theorem \cite{RS2}. The resolvent of $H$ could be found in two steps: First, we apply the resolvent of the free part to the equation (\ref{inhomogenous eq}) 
\begin{equation}
(H_0-z)^{-1} \rho = \psi - \sum_{i=1}^{N} \lambda_i \langle \varphi_i, \psi \rangle \; (H_0-z)^{-1} \varphi_i \;,
\end{equation}
and project this on the vector $\varphi_j$, we can then find the solution $\langle \varphi_i, \psi \rangle$ so that the resolvent $R_z(H)=(H-z)^{-1}$ of the Hamiltonian $H$ at $z$ is:
\begin{eqnarray}
R_z(H)=R_z(H_0) + \sum_{i,j=1}^{N} R_z(H_0) \varphi_i \; [\Phi^{-1}(z)]_{ij} \; \langle R_{\bar{z}}(H_0) \varphi_j, . \rangle \;, \label{Krein1}
\end{eqnarray}
where
\begin{eqnarray}
\Phi_{ij}(z)= \left\{ \begin{array} { l l } { {1 \over \lambda_i} - \langle \varphi_i, R_z(H_0) \varphi_i \rangle} & { \text { if } i = j } \\ { - \langle \varphi_i, R_z(H_0) \varphi_j \rangle}  & { \text { if } j \neq j' }\end{array} \right. \;. \label{Phi_finiterankperturbation}
\end{eqnarray}
Actually, the resolvent formula (\ref{Krein1}) is valid even in the case where the vectors $\varphi_i$'s do not belong to the Hilbert space. Such perturbations represent the singular type of interactions, e.g. Dirac delta potentials supported by points or curves \cite{Albeverio, ExnerKondej1}. In Dirac's bra-ket notation, one can also express the above resolvent formula as:
\begin{eqnarray}
R_z(H)=R_z(H_0) + \sum_{i,j=1}^{N} R_z(H_0) |\varphi_i \rangle \; [\Phi^{-1}(z)]_{ij} \; \langle \varphi_j | R_z(H_0)  \;. \label{Krein2}
\end{eqnarray}
The explicit expression of the resolvent (\ref{Krein1}) or (\ref{Krein2}) is known as Krein's resolvent formula. Alternatively, these singular interactions can be  defined directly  through von Neumann's self-adjoint extension theory (or quadratic forms in some cases). Since our aim is the spectral behaviour and especially the bound state problem of such singular interactions, Krein's explicit formula is much more useful. Throughout the paper, following the terminology introduced by S. G. Rajeev in \cite{Rajeev} we  call the matrix $\Phi$ as the principal matrix (this is equivalent to the matrix $\Gamma$ used in \cite{Albeverio}).

Actually,  one can also develop the above resolvent formula (\ref{Krein2}) to  relativistic  and field theoretical extensions of the singular models, as we will discuss in the next subsections. Let us now summarize explicitly the resolvent formulae and principal matrices in all classes of singular interactions that we are going to discuss in this paper:

\subsection{Point-like Dirac delta interactions in $\mathbb{R}$}

The Hamiltonian for the non-relativistic particle moving in fixed $N$ point like Dirac delta potentials in one dimension can be expressed in terms of the formal projection operators given by the Dirac kets $|a_i \rangle$
\begin{equation}
H=H_0 - \sum_{i=1}^{N} \lambda_i |a_i \rangle \langle a_i | \;, \label{Hamiltonian delta 1D}
\end{equation}
where $H_0$ is the non-relativistic free Hamiltonian, and $\lambda_j$'s are positive constants, called coupling constants or strengths of the potential. 
Throughout this paper, {\it we will use the units such that $\hbar=2m=1$ for non-relativistic cases and $\hbar=c=1$ only for the relativistic case}. Since we have fairly complicated expressions, this simplifies our writing, hoping that this does not lead to any further complications.  It is well-known in the literature that there are different ways to make sense of this formal Hamiltonian  in a mathematically rigorous way (strictly speaking, the above expression (\ref{Hamiltonian delta 1D}) has no meaning as an operator in $L^2(\mathbb{R})$). 
Let us define 
$R_z(H): =  R(z)$ and $R_z(H_0): = R_0(z)$ 
for simplicity. Even though it is hard  to make sense of the Hamiltonian, one can find the resolvent of this formal  operator algebraically  and the result is consistent with the one given by a  more rigorous formulation. 
Choosing $\varphi_i$ as the Dirac kets $|a_i\rangle$ formally in the previous section, the resolvent is explicitly given  by  
\begin{equation}
R(z)=R_0(z) + \sum_{i,j=1}^{N} R_0(z) |a_i \rangle [\Phi^{-1}(E)]_{ij} \langle a_{j}| R_0(z) \;, \label{resolvent delta 1D}
\end{equation}
where $\Phi$ is an $N \times N$ matrix
\begin{eqnarray}
\Phi_{i j}(z)= \left\{ \begin{array} { l l } { {1 \over \lambda_i} - R_0(a_i,a_i;z) } & { \text { if } i = j } \\ { - R_0(a_i,a_{j};z) } & { \text { if } i \neq j }\end{array} \right. \;.
\end{eqnarray}
Here $R_0(a_i,a_{j};z) = \langle a_i | (H_0-z)^{-1}|a_{j}\rangle$ is the free resolvent kernel. It is useful to express the principal matrix in terms of the heat kernel $K_t(a_i,a_j)$ - the fundamental solution to the Cauchy problem associated with the heat equation - using 
\begin{eqnarray}
R_0(a_i,a_{j};z)= \langle a_i | (H_0-z)^{-1}|a_{j}\rangle= \langle a_i | \int_{0}^{\infty} dt \, e^{t(H_0-z)} | a_{j}\rangle = \int_{0}^{\infty} dt \; K_t(a_i, a_j) \, e^{t z} \;.
\end{eqnarray} 
Then, we obtain
\begin{eqnarray}
\Phi_{i j}(z)= \left\{ \begin{array} { l l } { {1 \over \lambda_i} - \int_{0}^{\infty} dt \, K_t(a_i, a_i) e^{t z} } & { \text { if } i = j } \\ { - \int_{0}^{\infty} dt \, K_t(a_i, a_j) e^{t z} } & { \text { if } i \neq j }\end{array} \right. \;. \label{Phi_heatkernel1D}
\end{eqnarray}
These expressions should be considered as analytical continuations of the formulae beyond their regions of convergence in the variable $z$.
From the resolvent (\ref{resolvent delta 1D}), one can also write down the resolvent kernel
\begin{equation}
R(x_1,x_2;z)=R_0(x_1,x_2;z) + \sum_{i,j=1}^{N} R_0(x_1,a_i;z) [\Phi^{-1}]_{ij} R_0(a_{j},x_2;z) \;.
\end{equation}
Using the explicit expression of the integral kernel of the free resolvent 
\begin{equation}
R_0(x,y;z)= {i \over 2 \sqrt{z}} \; e^{i \sqrt{z}|x-y|} \;,
\end{equation}
we have
\begin{equation}
\Phi _ {i j} \left(z\right) =  \left\{ \begin{array} { l l } { {1 \over \lambda_i} - {i \over 2 \sqrt{z}}} & { \text { if } i = j } \\ { -{i \over 2 \sqrt{z}} \; e^{i \sqrt{z} |a_i-a_{j}|}} & { \text { if } i \neq j } \end{array} \right. \;. \label{principal matrix 1D z}
\end{equation} 
Here $\sqrt{z}$ is defined as the unambiguous square root of $z$ with $\Im{\sqrt{z}}$ is positive. Since we study the bound state spectrum, it is sometimes convenient to express the above matrix $\Phi(z)$ in terms of a real positive variable $\nu=-i \sqrt{z}$, i.e., 
\begin{equation}
\Phi_{ij}(z)\bigg|_{z=-\nu^2}:= \Phi_{ij}(\nu) = \left\{ \begin{array} { l l } { {1 \over \lambda_i} - {1 \over 2 \nu}} & { \text { if } i = j } \\ { -{1 \over 2 \nu} \; e^{-\nu |a_i-a_{j}|}} & { \text { if } i \neq j } \end{array} \right. \;. \label{Phi 1D}
\end{equation}

\subsection{Point-like Dirac delta Interactions in $\mathbb{R}^2$ and $\mathbb{R}^3$}

We assume that the centers of the Dirac delta potentials do not coincide, that is, $\mathbf{a}_i \neq \mathbf{a}_j$ whenever $i \neq j$. If we follow the same steps outlined above, we find exactly the same formal expression for the resolvent for point interactions in two and three dimensions except for the fact that the explicit expression of the integral kernel of the free resolvent in $\mathbb{R}^2$ and $\mathbb{R}^3$ \cite{Albeverio} are given by
\begin{eqnarray}
R_0(\mathbf{r_1},\mathbf{r_2};z)&=&{i \over 4} \; H_{0}^{(1)}(\sqrt{z}|\mathbf{r_1}-\mathbf{r_2}|) \;,
\\
R_0(\mathbf{r_1},\mathbf{r_2};z) & = &    {e^{i \sqrt{z} |\mathbf{r_1} -\mathbf{r_2}|} \over 4 \pi |\mathbf{r_1} -\mathbf{r_2}|} \;,
\end{eqnarray}
respectively. Here $H_{0}^{(1)}$ is the Hankel function of the first kind of order zero and $\Im{\sqrt{z}}>0$. Unfortunately, the diagonal part of the free resolvent kernels are divergent so the diagonal part of the principal matrices are infinite. This is clear for the three dimensional case from the asymptotic behavior of the Hankel function \cite{Lebedev}
\begin{equation}
H_{0}^{(1)}(x) \approx -{2i \over \pi} \log(2/x) \;,
\end{equation}
as $x\rightarrow 0$.

 This difficulty  can be resolved by the so-called  regularization and renormalization method. Instead of starting with the higher dimensional version of the formal Hamiltonian (\ref{Hamiltonian delta 1D}),
we first consider the regularized Hamiltonian through the heat kernel 
\begin{equation} \label{regularizedHpointflat}
H_{\epsilon}=H_0 - \sum_{i=1}^{N} \lambda_i (\epsilon) \; |\mathbf{a}_{i}^{\epsilon}   \rangle \langle \mathbf{a}_{i}^{\epsilon}| \;,
\end{equation}
where $\langle \mathbf{r}| \mathbf{a_{i}}^{\epsilon} \rangle = K_{\epsilon/2}(\mathbf{r},\mathbf{a_i})$. 
The heat kernel associated with the heat equation $\nabla^2 \psi - {\partial \psi \over \partial t} =0$ in $\mathbb{R}^n$ is given by
\begin{equation}
K_t(\mathbf{r_1},\mathbf{r_2})={1 \over (4 \pi t)^{n/2}} e^{-{|\mathbf{r_1}-\mathbf{r_2}|^2 \over 4 t} }\;. \label{heatkernelRn}
\end{equation}
It is important to note that 
\begin{equation} \label{limit of heat kernel}
K_{\epsilon/2}(\mathbf{r},\mathbf{a_i}) \rightarrow \delta(\mathbf{r}-\mathbf{a_i}) \;,
\end{equation}
as $\epsilon \rightarrow 0^{+}$ in the distributional sense. Then, we can easily find the resolvent kernel associated with the regularized Hamiltonian (\ref{regularizedHpointflat})
\begin{equation}
R_{\epsilon}(\mathbf{r_1},\mathbf{r_2};z) = R _ {0}( \mathbf{r_1},\mathbf{r_2};z) + \sum_{i,j = 1}^{N} R _ {0} \left(\mathbf{r_1},\mathbf{a_{i}} ; z \right) 
\left[\Phi_{\epsilon}(z)\right]^{-1}_{ij} R _ { 0 } \left(\mathbf{a_{j}},\mathbf{r_2} ; z \right) \;,
\end{equation}
where 
\begin{equation}
[\Phi_{\epsilon}(z)]_{i j} = \left\{ \begin{array} { l l } { {1 \over \lambda_i(\epsilon)} - \int_{0}^{\infty} d t \;K_{t+\epsilon}(\mathbf{a_i},\mathbf{a_i})\; e^{t z}} & { \text { if } i = j } \\ { -\int_{0}^{\infty} d t \;K_{t+\epsilon}(\mathbf{a_i},\mathbf{a_{j}}) \; e^{t z}} & { \text { if } i \neq j } \end{array} \right. \;.
\end{equation}
If we choose 
\begin{equation} \label{Coupling constant choice for renormalization}
{1 \over \lambda_i(\epsilon)} = \int_{0}^{\infty} d t \;K_{t+\epsilon}(\mathbf{a_i},\mathbf{a_i})\; e^{ t E_{B}^{i}}
\end{equation}
where $E_{B}^{i}<0$ (the spectrum of the free Hamiltonian only includes the continuous spectrum: $[0, \infty]$) is the bound state energy of the particle to the $i$ th center {\it in the absence of all the other centers} and take the formal limit $\epsilon \rightarrow 0^+$ we find
\begin{equation}
R(\mathbf{r_1},\mathbf{r_2};z) = R _ {0}( \mathbf{r_1},\mathbf{r_2};z) + \sum_{i,j = 1}^{N} R _ {0} \left(\mathbf{r_1},\mathbf{a_{i}} ; z \right) 
\left[\Phi(z)\right]^{-1}_{ij} R _ { 0 } \left(\mathbf{a_{j}},\mathbf{r_2} ; z \right) \;, \label{resolvent_point 2D 3D}
\end{equation}
where 
\begin{eqnarray}
\Phi_{i j}(z) & = & \left\{ \begin{array} { l l } {\int_{0}^{\infty} d t \; K_{t}(\mathbf{a_i},\mathbf{a_i}) \; \left(e^{t E_{B}^{i}} - e^{t z} \right)} & { \text { if } i = j } \\ { -\int_{0}^{\infty} d t \;K_{t}(\mathbf{a_i},\mathbf{a_{j}}) \; e^{t z}} & { \text { if } i \neq j } \end{array} \right. \;.\label{NR_point_Phi_heat_kernel}
\end{eqnarray}
From the explicit form of the heat kernel formula (\ref{heatkernelRn}), we obtain 
\begin{equation}
\Phi_{i j}(z) = \left\{ \begin{array} { l l } {{1 \over 2\pi} \log\left(-i\sqrt{z/|E_{B}^{i}|}\right)} & { \text { if } i = j } \\ { -{i \over 4} H_0^{(1)}(\sqrt{z} |\mathbf{a_i} -\mathbf{a_{j}}|)} & { \text { if } i \neq j } \end{array} \right. \;, \label{principal matrix in 2D}
\end{equation}
\textit{in two dimensions} and
\begin{equation} \label{principal matrix in 3D}
\Phi_{i j}(z) = \left\{ \begin{array} { l l } {\left(-i\sqrt{z}-\sqrt{|E_{B}^{i}|}\right) \over 4\pi} & { \text { if } i = j } \\ { -{e^{i \sqrt{z} |\mathbf{a_i} -\mathbf{a_{j}}|} \over 4 \pi |\mathbf{a_i} -\mathbf{a_{j}}|}} & { \text { if } i \neq j } \end{array} \right. \;,
\end{equation}
\textit{in three dimensions}.

Since we deal with the bound states in this paper, it is convenient to express the principal matrices in terms of the real positive variable $\nu=-i\sqrt{z}$:

\begin{equation}
\Phi_{i j}(z)|_{z=-\nu^2} = \left\{ \begin{array} { l l } {{1 \over 2\pi} \log \left(\nu/ \sqrt{|E_{B}^{i}|}\right)} & { \text { if } i = j } \\ { -{1 \over 2\pi} K_0(\nu |\mathbf{a_i} -\mathbf{a_{j}}|)} & { \text { if } i \neq j } \end{array} \right. \;, \label{principal matrix in 2D real}
\end{equation}
\textit{in two dimensions} and
\begin{equation} \label{principal matrix in 3D real}
\Phi_{i j}(z)|_{z=-\nu^2} = \left\{ \begin{array} { l l } {\left(\nu -\sqrt{|E_{B}^{i}|}\right) \over 4\pi} & { \text { if } i = j } \\ { -{e^{-\nu |\mathbf{a_i} -\mathbf{a_{j}}|} \over 4 \pi |\mathbf{a_i} -\mathbf{a_{j}}|}} & { \text { if } i \neq j } \end{array} \right. \;,
\end{equation}
\textit{in three dimensions}. Here we have used $K_0(z)={i \pi \over 2} \, H_{0}^{1}(i z)$ with $-\pi<arg(z)<\pi/2$ and $K_0(z)$ is the modified Bessel function of the third kind \cite{Lebedev}.

\subsection{Point-like Dirac delta Interactions in $\mathbb{H}^2$ and $\mathbb{H}^3$}

Here we assume that the particle is intrinsically moving in the manifold.
Our heuristic approach to study such type of interactions on Riemannian manifolds is based on the idea of using the heat kernel as a regulator for point interactions on manifolds  \cite{AltunkaynakErmanTurgut, ErmanTurgut}. Thanks to the fact (\ref{limit of heat kernel}), the regularized interaction is chosen as the heat kernel on Riemannian manifolds. Once we have regularized the Hamiltonian, one can follow essentially the same steps outlined in the previous section, and obtain exactly the same form of the resolvent and principal matrix as in (\ref{resolvent_point 2D 3D}) and (\ref{NR_point_Phi_heat_kernel}), respectively. In this paper, we only consider the particular class of Riemannian manifolds, namely two and three dimensional hyperbolic manifolds for simplicity. The heat kernel on hyperbolic manifolds of constant sectional curvature $-\kappa^2$ can be analytically calculated and given by \cite{Grigoryan Heat Book}
\begin{eqnarray}
 \label{heatkernelhperbolicspaces} 
K_{t}(x,y) =
\begin{cases}
\begin{aligned}
{\sqrt{2} \over \kappa} {1 \over (4 \pi t)^{3/2}} \; e^{- \kappa^2 t/4} \int_{\kappa  d(x,y)}^{\infty} d s \; 
{s \; e^{-s^2/4 \kappa^2 t} \over \sqrt{\cosh s -\cosh \kappa d(x,y)}}   \;
\end{aligned}
& \mathrm{for} \; n=2 \\[2ex]
\begin{aligned}
 { \kappa d(x,y)  \over (4 \pi t)^{3/2} \sinh  \kappa d(x,y)} \; e^{- \kappa^2 t -{d^2(x,y) \over 4 t}} \;
\end{aligned}
& \mathrm{for} \; n=3 \;,
\end{cases}
\end{eqnarray}
where $d(x,y)$ is the geodesic distance between the points $x$ and $y$ on the manifold. The explicit form of the principal matrix in $\mathbb{H}^3$ can then be easily evaluated \cite{ErmannumberH}:
\begin{eqnarray}
\Phi_{ij}(z)= 
\begin{cases}
\begin{aligned}
{1 \over 4 \pi} \left(\sqrt{\kappa^2 -z}-\sqrt{\kappa^2- E_{B}^{i}}\right)
\end{aligned}
& \mathrm{if} \; i=j \\[2ex]
\begin{aligned}
- \left( {\kappa \exp \left(- d(a_i,a_{j})  \sqrt{\kappa^2 -z}\right) \over 4 \pi \sinh \left(\kappa d(a_i,a_{j}) \right) } \right)  \;
\end{aligned}
& \mathrm{if} \; i \neq j \;.
\end{cases}
 \label{PhiH3}
\end{eqnarray}
Similarly, the principal matrix in $\mathbb{H}^2$ can simply be evaluated by interchanging the order of integration with respect to $t$ and $s$ 
\begin{eqnarray}
\Phi_{ij}(z)  =
\begin{cases}
\begin{aligned}
 & {1 \over 2 \pi} \left[ \psi \left( {1 \over 2} + \sqrt{-{z \over \kappa^2} + {1 \over 4}} \right)  - \psi \left( {1 \over 2} + \sqrt{-{E_{B}^{i} \over \kappa^2} +{1 \over 4}} \right) \right]    
\end{aligned}
& \mathrm{if} \; i=j \\
\begin{aligned} 
- {1 \over 2 \pi} \; Q_{ {1 \over 2} + \sqrt{-{z \over \kappa^2} + {1 \over 4}} } \left(\cosh(\kappa d(a_i,a_j) )\right) 
\end{aligned}
& \mathrm{if} \; i \neq j \;,
\end{cases}
\label{PhiH2} 
\end{eqnarray}
where $\psi$ is the digamma function with its integral representation  \cite{Lebedev}
\begin{equation}
\psi(w)=\int_{0}^{\infty} \left( {e^{-t} \over t} -{e^{- t w} \over 1-e^{-t}} \right) \; d t \;, \label{intreppsi}
\end{equation}
for $\Re(w)>0$, and $Q$ is the Legendre function of second type \cite{Lebedev} with its integral representation
\begin{equation}
Q_{\alpha}(\cosh a)=\int_{a}^{\infty} {e^{-(\alpha + {1 \over 2}) r} \over \sqrt{2 \cosh r -2 \cosh a}} \; d r \;, \label{intrepQ}
\end{equation}
for real and positive $a$ and $\Re(\alpha)>-1$.

Since the spectrum of the free Hamiltonian in $\mathbb{H}^{n}$ includes only the continuous part starting from $(n-1)^2 \kappa^2/4$, it is natural to assume $E_{B}^{i}<(n-1)^2\kappa^2/4$.

\subsection{Two Types of Relativistic Extensions of Point-like Dirac delta Interactions}

We first consider the so-called semi-relativistic Salpeter type free Hamiltonian (also known as relativistic spin zero Hamiltonian) perturbed by point like Dirac delta potentials in one dimension. This problem for the single center case has been first studied in \cite{Albeverio Kurasov Salpeter} from the self-adjoint extension point of view. The formal Hamiltonian is exactly in the same form as in (\ref{Hamiltonian delta 1D}), except for the free part 
\begin{equation}
H= \sqrt{-{d^2 \over dx^2} +m^2} - \sum_{i=1}^{N} \lambda_i |a_i \rangle \langle a_i| \;,
\end{equation}
in the units where $\hbar=c=1$. This non-local operator is a particular  pseudo-differential operator and defined in momentum
space as multiplication by $\sqrt{p^2 +m^2}$ \cite{LiebLoss}, which is known as the symbol of the operator. After following the renormalization procedure outlined above for the point interactions in two and three dimensions, the resolvent and the principal matrix is exactly the same form as in (\ref{resolvent_point 2D 3D}) and (\ref{NR_point_Phi_heat_kernel}), respectively. However, the explicit expression of the heat kernel in this case is given by \cite{LiebLoss}
\begin{equation}
K _ { t } ( x , y ) = \frac { m t } { \pi \sqrt { ( x - y ) ^ { 2 } + t ^ { 2 } } } K _ { 1 } \left( m \sqrt { ( x - y ) ^ { 2 } + t ^ { 2 } } \right)  \;, \label{heat kernel salpater} \end{equation}
where $K_{1}$ is the modified Bessel function of the first kind. Due to the short-time asymptotic expansion 
\begin{equation}
K_1(mt) \sim {1 \over m t}     \;,
\end{equation}
the diagonal term in the principal matrix (\ref{NR_point_Phi_heat_kernel}) is divergent. In contrast to the one-dimensional case for point Dirac delta potentials, this problem therefore requires renormalization, as noticed by \cite{ermangadellauncu, Alhashimi}. Choosing the coupling constants as in (\ref{Coupling constant choice for renormalization}) by substituting the heat kernel (\ref{heat kernel salpater}) and taking the limit $\epsilon \rightarrow 0^+$, we obtain the resolvent in the form of the Krein's formula (\ref{resolvent delta 1D}). The explicit form of the diagonal principal matrix is given by \cite{ermangadellauncu}
\begin{equation}
\Phi_ {ii} (z) = \varphi(E_{B}^i) - \varphi(z) \;,
\end{equation}
where
\begin{equation}
\varphi(z)=\frac { z } { \pi \sqrt { m ^ { 2 } - z ^ { 2 } } } \left( \frac { \pi } { 2 } + \arctan \frac { z } { \sqrt { m ^ { 2 } - z ^ { 2 } } } \right) \;.
\end{equation}
Its off-diagonal part is given by
\begin{equation}
\Phi_ { i j } ( z ) = \left\{ \begin{array} { l l } { - \frac { 1 } { \pi } \int _ { m } ^ { \infty } d \mu \; e ^ { - \mu \left| a _ { i } - a _ { j } \right| } \frac { \sqrt { \mu ^ { 2 } - m ^ { 2 } } } { \mu ^ { 2 } - m ^ { 2 } + z ^ { 2 } } } & { \text { if }  \Re{z}  < 0 } \\ { - i \frac { e ^ { i \sqrt { z ^ { 2 } - m ^ { 2 } } \left| a _ { i} - a _ { j } \right| } } { \sqrt { 1 - \frac { m ^ { 2 } } { z ^ { 2 } } } } - \frac { 1 } { \pi } \int _ { m } ^ { \infty } d \mu \; e ^ { - \mu \left| a _ { i } - a _ { j } \right| } \frac { \sqrt { \mu ^ { 2 } - m ^ { 2 } } } { \mu ^ { 2 } - m ^ { 2 } + z ^ { 2 } } } & { \text { if }  \Re{z}  > 0 } \end{array} \right. \;, \label{relativistic Phi}
\end{equation}
where $E_{B}^i$ is the bound state energy to the $i$ th center in the absence of all the other centers. Since the spectrum of the free Hamiltonian includes only the continuous spectrum starting from $m$, it is natural to expect that $E_{B}^{i}<m$.

An alternative relativistic model can be introduced from a field theory perspective in two dimensions. If we take very heavy particles interacting with  a light particle, in the extreme  limit of {\it static} heavy particles one recovers the following model:
\begin{equation}
H=\iint_{\mathbb{R}^2} {d^2 \mathbf{p} \over (2 \pi)^2} \;   \sqrt{\left(\mathbf{p}^2+m^2 \right)} \; a^\dagger(\mathbf{p}) a(\mathbf{p})  - \sum_{i=1}^{N} \lambda_i\phi^{(-)} (\mathbf{a_i})\phi^{(+)} (\mathbf{a_i})
\;,\end{equation}
where $\mathbf{a_i}$ refer to the locations of static heavy particles. Here 
\begin{equation}
    \phi^{(+)}(\mathbf{x})=\iint_{\mathbb{R}^2} {d^2 \mathbf{p} \over (2 \pi)^2}\; {e^{i \mathbf{p} \cdot \mathbf{x}} \over \sqrt{2} (\mathbf{p}^2+m^2)^{1/4}} a(\mathbf{p}) \quad {\rm and}\ \ \phi^{(-)}=\big(\phi^{(+)} \big)^\dagger \;,
\end{equation}
where $\dagger$ denotes the adjoint. Since this model was worked out in \cite{Caglar1}, we will be content with the resulting formulae only referring to the original paper for the details. 
We can compute the diagonal principal matrix as
\begin{eqnarray}
\Phi_{ii}(z)= {1\over 2\pi} \ln \Big({m-z\over m-E^i_B}\Big) \;,
\end{eqnarray}
and the off-diagonal part as
\begin{eqnarray}
\Phi_{ij}(z)=-{1\over 2\pi}\int_0^\infty {ds \over (s^2+1)^{1/2}} \; e^{-|\mathbf{a_i}-\mathbf{a_j}| [m(s^2+1)^{1/2} -z s]} \;,
\label{Phi relativistic point 2}
\end{eqnarray}
for $-m<\Re{(z)}<m$. Moreover, the binding energy of the single center should satisfy  $-m< E_B^{i}<m$, and the lower bound is due to the stability requirement, to prevent pair creation to reduce the energy further thus rendering the model unrealistic in single particle sector.

\subsection{Dirac delta Interactions supported by curves in $\mathbb{R}^2$ and in $\mathbb{R}^3$}

We consider $N$ Dirac delta potentials supported by non-intersecting smooth curves $\mathbf{\gamma_j}:[0,L_j] \rightarrow \mathbb{R}^n$ of finite length $L_j$ ($n=2,3$). Each curve is assumed to be simple, i.e., $\mathbf{\gamma_j}(s_1) \neq \mathbf{\gamma_j}(s_2)$  whenever $s_1 \neq s_2$, where $s_1, s_2 \in (0,L_j)$. Our formulation also allows the simple closed curves. 

The Hamiltonian of the system is given by
\begin{equation}
H= H_0 - \sum_{i=1}^{N} {\lambda_i \over L_i} | \mathbf{\gamma_i} \rangle \langle \mathbf{\gamma_i} | \;,
\end{equation}
where $\langle   \mathbf{\gamma_i}  | \mathbf{r} \rangle=\int_{\Gamma_i} d s \; \delta(\mathbf{r}-\mathbf{\gamma_i}(s)) $. Then, the Schr\"{o}dinger equation ($H |\psi \rangle = E |\psi \rangle$) associated with this Hamiltonian is 
\begin{eqnarray}
- \nabla^{2} \psi(\mathbf{r}) - \sum_{i=1}^{N} {\lambda_i \over L_i} \int_{\Gamma_i} d s_i \; \delta(\mathbf{r}-\mathbf{\gamma_i}(s_i)) \; \int_{\Gamma_i} d s'_{i} \; \psi(\mathbf{\gamma_i}(s'_{i})) = E \psi(\mathbf{r}) \;.
\end{eqnarray}
In contrast to the point-like Dirac delta interactions, this equation is a generalized Schr\"{o}dinger equation in the sense that it is non-local. The resolvent kernel of the above Hamiltonian is explicitly given in the same form associated with point like Dirac delta potentials, namely  
\begin{equation}
R(\mathbf{r_1},\mathbf{r_2};z) = R _ {0}( \mathbf{r_1},\mathbf{r_2};z) +  \sum_{i,j = 1}^{N} {1 \over \sqrt{L_i L_{j}}} R _ {0} \left(\mathbf{r_1},\mathbf{\gamma_{i}} ; z \right) 
\left[\Phi(z)\right]^{-1}_{ij} R _ { 0 } \left(\mathbf{\gamma_{j}},\mathbf{r_2} ; z \right) \;, \label{resolventkernelcurveR2}
\end{equation}
where
\begin{equation} \label{principal matrix resolvent curve 2D}
\Phi_{i j}(z) = \left\{ \begin{array} { l l } {{1 \over \lambda_i}- {1 \over L_i} \langle \mathbf{\gamma_i} | R_0(z) | \mathbf{\gamma_i} \rangle } & { \text { if } i = j } \\ { -{1 \over \sqrt{L_i L_{j}}} \langle \mathbf{\gamma_i} | R_0(z) | \mathbf{\gamma_{j}} \rangle } & { \text { if } i \neq j } \end{array} \right. \;,
\end{equation}
or if we express it in terms of the heat kernel 
\begin{equation}
\Phi_{i j}(z) = \left\{ \begin{array} { l l } { {1 \over \lambda_i} - {1 \over L_i} \iint_{\Gamma_i \times \Gamma_{i}} ds_i \; ds'_{i} \;  \int_{0}^{\infty} dt \;e^{t z} \; K_{t}(\mathbf{\gamma_i}(s_i),\mathbf{\gamma_i}(s'_{i}))} & { \text { if } i = j } \\ { - {1 \over \sqrt{L_i L_{j}}} \iint_{\Gamma_i \times \Gamma_{j}} ds_i \; ds'_{j} \;  \int_{0}^{\infty} dt \;e^{t z} \; K_{t}(\mathbf{\gamma_i}(s_j),\mathbf{\gamma_{j}}(s'_{j})) } & { \text { if } i \neq j } \end{array} \right. \;.
\label{Phi_heatkernel_curves}\end{equation}
Using the explicit form of the heat kernel in two dimensions, the above principal matrix becomes
\begin{equation}
\Phi_{i j}(z) = \left\{ \begin{array} { l l } { {1 \over \lambda_i} - {i \over 8 \pi L_i} \iint_{\Gamma_i \times \Gamma_{i}} ds_i \; ds'_{i} \;   H_{0}^{(1)}(\sqrt{z} |\mathbf{\gamma_i}(s_i) - \mathbf{\gamma_i}(s'_{i})|)} & { \text { if } i = j } \\ { - {i \over 8 \pi \sqrt{L_i L_{j}}} \iint_{\Gamma_i \times \Gamma_{j}} ds_i \; ds'_{j} \;  H_{0}^{(1)}(\sqrt{z} |\mathbf{\gamma_i}(s_i)-\mathbf{\gamma_{j}}(s'_{j})|) } & { \text { if } i \neq j } \end{array} \right. \;. \label{Phi_curve2D}
\end{equation}
The spectrum of the free Hamiltonian includes only continuous spectrum starting from zero, so we expect that the bound state energies must be below $z=0$. For this reason, we restrict the principal matrix to the negative real values, i.e., $z=-\nu^2$, $\nu>0$. Then, we have
\begin{equation}
\Phi_{i j}(z)|_{z=-\nu^2} = \left\{ \begin{array} { l l } { {1 \over \lambda_i} - {1 \over 4 \pi L_i} \iint_{\Gamma_i \times \Gamma_{i}} ds_i \; ds'_{i} \;   K_{0}(\nu |\mathbf{\gamma_i}(s_i) - \mathbf{\gamma_i}(s'_{i})|)} & { \text { if } i = j } \\ { - {1 \over 4 \pi \sqrt{L_i L_{j}}} \iint_{\Gamma_i \times \Gamma_{j}} ds_i \; ds'_{j} \;  K_{0}(\nu |\mathbf{\gamma_i}(s_i)-\mathbf{\gamma_{j}}(s'_{j})|) } & { \text { if } i \neq j } \end{array} \right. \;. \label{Phi_curve2D real}
\end{equation}
For non self-intersecting curve $\gamma_i$, we can expand it around the neighbourhood of $s'_{i}=s_i$ in the Serret-Frenet frame at $s_i$ \cite{DoCarmo}:
\begin{equation}
\mathbf{\gamma_i}(s'_{i}) = \mathbf{\gamma_i}(s_i) + \left( (s'_{i}-s_i)-k_{i}^2(s_i){(s'_{i}-s_i)^3 \over 3!}\right) \mathbf{t}_i(s_i)+ \left({k_i(s_i) \over 2} (s'_{i}-s_i)^2 -k_{i}^{'}(s_i){(s'_{i}-s_i)^3 \over 3!} \right) \mathbf{n}_i(s_i) + \mathbf{R}_i(s_i) \;,
\end{equation} 
where $\mathbf{t}_i(s_i)$ and $\mathbf{n}_i(s_i)$ are the tangent and normal vectors at $s_i$, and $\mathbf{R}_i(s_i)$ is the remainder term which vanishes faster than $(s'_{i}-s_i)^3$ as $s'_{i}\rightarrow s_i$. We have an extra term proportional to the binormal vector $\mathbf{b}_i(s_i)$ in three dimensions ($-{k_i(s_i) \tau_i(s_i) \over 3!}(s'_{i}-s_i)^3 \mathbf{b}_i(s_i)$, where $\tau_i(s_i)$ is the torsion of the curve). In the first approximation, keeping only the linear terms in $s'_{i}-s_i$, and translating and rotating the Serret-Frenet frame attached to the coordinate system $Oxy$ in such a way that $\mathbf{t}_i(s_i)=(1,0)$ and $\mathbf{n}_i(s_i)=(0,1)$, we have
\begin{equation}
|\mathbf{\gamma_i}(s'_{i}) - \mathbf{\gamma}_i(s_i)| \approx | s'_{i}-s_i | \;. 
\end{equation} 
Then, the integral in the diagonal part of the principal matrix (\ref{Phi_curve2D real}) around $s'_{i}=s_i$ in the first approximation is
\begin{equation}
\iint_{\Gamma_i \times \Gamma_{i}} ds_i \; ds'_{i} \;   K_{0}(\nu | s'_{i}-s_i |) \;.
\end{equation}
By making change of coordinates $\xi_i={ (s'_{i}+s_i) \over 2}$ and $\eta_i= {(s'_{i}-s_i) \over 2}$, the above integral becomes
\begin{equation}
4 \int_{0}^{L_i/2} d \eta_i (L_i- 2 \eta_i) K_0(2 \nu \eta_i) \;.
\end{equation}
Using $\int_{0}^{L_i/2} d \eta_i (L_i- 2 \eta_i) K_0(2 \nu \eta_i) \leq \int_{0}^{\infty} d \eta_i (L_i- 2 \eta_i) K_0(2 \nu \eta_i)$ and the integrals of modified Bessel functions \cite{Table} 
\begin{equation}
\int_{0}^{\infty} d x \; x^{n} \, K_{0}(a x) = 2^{n-1} a^{-n-1} \Gamma^2 \left({1+n \over 2}\right) \;,
\end{equation}
where $n=0,1$ and $\Gamma$ is the gamma function, it is easy to see that the integral that we consider is finite around $\eta_i=0$ ($s'_{i}=s_i$). For non self-intersecting curves, the integrals in the diagonal and off-diagonal terms in (\ref{Phi_curve2D real}) are finite whenever $s'_{i} \neq s_i$ due to the upper bounds of the Bessel functions \cite{ErmanTurgut} 
\begin{equation}
K _ { 0 } ( x ) < \frac { 2 } { 1 + x } \mathrm { e } ^ { - \frac { x } { 2 } } + \mathrm { e } ^ { - \frac { x } { 2 } } \ln \left( \frac { x + 1 } { x } \right) \;.
\end{equation}

In three dimensions, the Dirac delta potentials supported by curves requires the renormalization. Using the explicit formula of the heat kernel (\ref{heatkernelRn}) for three dimensions, we find  
\begin{equation}
\Phi_{i j}(z) = \left\{ \begin{array} { l l } { {1 \over \lambda_i} - {1 \over 4 \pi L_i} \iint_{\Gamma_i \times \Gamma_{i}} ds_i \; ds'_{i} \;  {e^{i \sqrt{z} |\mathbf{\gamma_i(s_i)}-\mathbf{\gamma_i(s'_{i})}|} \over |\mathbf{\gamma_i}(s_i)-\mathbf{\gamma_i}(s'_{i})|} } & { \text { if } i = j } \\ { - {1 \over 4 \pi \sqrt{L_i L_{j}}} \iint_{\Gamma_i \times \Gamma_{j}} ds_i \; ds'_{j} \;  {e^{i \sqrt{z} |\mathbf{\gamma_i(s_i)}-\mathbf{\gamma_j(s'_{j})}|} \over |\mathbf{\gamma_i}(s_i)-\mathbf{\gamma_j}(s'_{j})|} } & { \text { if } i \neq j } \end{array} \right. \;.
\label{Phi_curves_3D}\end{equation}
One can show that the the diagonal part of the above principal matrix (\ref{Phi_heatkernel_curves}) includes a term
\begin{eqnarray}
 \iint_{\Gamma_i \times \Gamma_{i}} ds_i \; ds'_{i} \; {e^{i \sqrt{z} |\mathbf{\gamma_i(s_i)}-\mathbf{\gamma_i(s'_{i})}|} \over |\mathbf{\gamma_i}(s_i)-\mathbf{\gamma_i}(s'_{i})|} \;,
\end{eqnarray}
which is  divergent around $s'_{i}=s_i$. This can be immediately seen using the similar method outlined above, that is, the above integral includes the following integral in the new variable $\eta_i$:
\begin{eqnarray}
\int_{0}^{L_i/2} d\eta_i \; {e^{2 i \sqrt{z} \eta_i} \over \eta_i} \;,
\end{eqnarray}
which is divergent around $\eta_i=0$. 

Similar to the non-relativistic and relativistic point interactions, we  first regularize the resolvent and then by choosing the coupling constant as a function of the cut-off parameter $\epsilon$:
\begin{equation}
{1 \over \lambda_i(\epsilon)} = \int_{0}^{\infty} dt \; e^{t E_{B}^{i}} K_{t+\epsilon}(\mathbf{\gamma_i}(s_i),\mathbf{\gamma_i}(s'_{i})) \;,
\end{equation}
and taking the formal limit $\epsilon \rightarrow 0^+$, we obtain the resolvent
which is exactly the same form as in (\ref{resolventkernelcurveR2}) except the matrix $\Phi$ is given by
\begin{equation}
\Phi_{i j}(z) = \left\{ \begin{array} { l l } { {1 \over L_i} \iint_{\Gamma_i \times \Gamma_{i}} ds_i \; ds'_{i} \;  \int_{0}^{\infty} dt \; (e^{t E_{B}^{i}}-e^{t z}) K_{t}(\mathbf{\gamma_i}(s_i),\mathbf{\gamma_i}(s'_{i}))} & { \text { if } i = j } \\ { -{1 \over \sqrt{L_i L_{j}}} \iint_{\Gamma_i \times \Gamma_{j}} ds_i \; ds'_{j} \;  \int_{0}^{\infty} dt \;e^{t z} \; K_{t}(\mathbf{\gamma_i}(s_i),\mathbf{\gamma_{j}}(s'_{j}))  } & { \text { if } i \neq j } \end{array} \right. \;.
\end{equation}
Here, $E_{B}^{i}$ is the bound state energy of the particle to the delta interaction supported by $i$th curve in the absence of all the other delta interactions. Since the spectrum of the free Hamiltonian only includes the continuous part starting from zero, we have $E_{B}^{i}<0$. Using the explicit form of the heat kernel, the principal matrix turns out to be a finite expression:
\begin{equation}
\Phi_{i j}(z) = \left\{ \begin{array} { l l } { {1 \over 4 \pi L_i} \iint_{\Gamma_i \times \Gamma_{i}} ds_i \; ds'_{i} \; {1 \over |(\mathbf{\gamma_i}(s_i)-\mathbf{\gamma_i}(s'_{i}))|}  (e^{-\sqrt{|E_{B}^{i}|}|(\mathbf{\gamma_i}(s_i)-\mathbf{\gamma_i}(s'_{i}))|}-e^{i\sqrt{z} |(\mathbf{\gamma_i}(s_i)-\mathbf{\gamma_i}(s'_{i}))|})} & { \text { if } i = j } \\ { -{1 \over 4 \pi \sqrt{L_i L_{j}}} \iint_{\Gamma_i \times \Gamma_{j}} ds_i \; ds'_{j} \; {e^{i \sqrt{z} |(\mathbf{\gamma_i}(s_i)-\mathbf{\gamma_{j}}(s'_{j}))|} \over |(\mathbf{\gamma_i}(s_i)-\mathbf{\gamma_{j}}(s'_{j}))|} \;  } & { \text { if } i \neq j } \end{array} \right. \;. \label{principal_curve_3D}
\end{equation}
A semi-relativistic generalization of particles interacting with  curves  is presented in \cite{BurakTeo}.
The formal Hamiltonian can be written as 
\begin{equation}
    H=\iint_{\mathbb{R}^2}  {d^2 \mathbf{p} \over (2 \pi)^2} \;  \left(\mathbf{p}^2+m^2 \right)^{1/2}  a^\dagger(\mathbf{p})  a(\mathbf{p}) - \sum_{i=1}^{N} {\lambda_i\over L_i}\int ds_i \; \phi^{(-)} (\gamma_i(s_i))\int ds'_i \; \phi^{(+)} (\gamma_i(s_i')).
\end{equation}
We refer to this work for the details and we are content with writing down the resulting $\Phi$ matrix, since for tunneling corrections to the bound spectra this is all we need:
\begin{eqnarray}
\Phi_{ii}(z)&=& {m\over \sqrt{2} \pi^2 L_i}\int_0^\infty dt \; \int_{\Gamma_i\times \Gamma_i}ds_i ds_i' \; {K_1\big(m\sqrt{t^2 +|\gamma_i(s_i)-\gamma_i(s_i')|^2}\big)\over \sqrt{t^2+|\gamma_i(s_i)-\gamma_i(s_i')|^2 }} \left(e^{E_B^i t} - e^{z t}\right) \;, \\
\Phi_{ij}(z) &=&-{m\over \sqrt{2L_iL_j} \pi^2 }\int_0^\infty dt \; \int_{\Gamma_i\times \Gamma_j}ds_i ds_j \;  {K_1\big(m\sqrt{t^2 +|\gamma_i(s_i)-\gamma_j(s_j)|^2}\big)\over \sqrt{t^2+|\gamma_i(s_i)-\gamma_j(s_j)|^2 }} e^{z t} \;.
\end{eqnarray}
As usual, these formulae must be analytically continued  in $z$ outside of their region of convergence. In our approach we are interested in the bound states for which these formulae are valid.

\section{Analytic Structure of the Principal Matrices and the Bound State Spectrum}
\label{Bound State Spectrum}

It is well-known that the bound state spectrum is determined by the poles of the resolvent, so the bound state spectrum should only come from the points $z$ below the spectrum of the free Hamiltonian, where the matrix $\Phi$ is not invertible, i.e., the bound state energies are the real solutions of the equation
\begin{equation}
\det \Phi(E)=0 \;,
\end{equation}
where $E < \sigma(H_0)$. Looking at the explicit forms of the principal matrices $\Phi_{ij}(z)$, we see that they are all matrix-valued holomorphic functions (defined on their largest possible set of the complex plane).
The analytical structure of the principal matrices can be determined by using the generalized Loewner's theorem \cite{Rosenblum Rovnyak}, which simply states that if $f_0$ is a real valued continuously differentiable function on an open subset $\Delta$ of $(-\infty, \infty)$, then the following are equivalent:
\begin{itemize}
    \item There exists a holomorphic function $f$ with $\Im{f} \geq 0$ on the upper half-plane of the complex plane such that $f$ has an analytic continuation  across $\Delta$ that coincides with $f_0$ on $\Delta$.
    
    \item For each continuous complex valued function $F$ on $\Delta$ that vanishes off a compact subset of $\Delta$,
    \begin{eqnarray}
    \int_{\Delta} \int_{\Delta} d\zeta\; d \eta \;  K(\zeta, \eta) \bar{F}(\zeta) F(\eta) \geq 0 \;,
    \end{eqnarray}
    where for $\zeta, \eta \in \Delta$,
    \begin{eqnarray}
    K(\zeta, \eta) = \left\{ \begin{array} { l l } { {f_0(\zeta)-f_0(\eta) \over \zeta - \eta}} & { \text { if } \zeta \neq \eta } \\ {f'_{0}(\zeta)} & { \text { if } \zeta = \eta } \end{array} \right. \;.
    \end{eqnarray}

    \end{itemize}
For simplicity, let us explicitly show the analytical structure of the principal matrix associated with the Dirac delta potential supported by a single curve in two dimensions. In this case, the principal matrix (\ref{principal matrix resolvent curve 2D}) is just the diagonal part, say $\Phi(E)$, and continuously differentiable function of $E$, where $E$ is on the negative real axis. Then, we have  
\begin{eqnarray}
{\Phi(\zeta)-\Phi(\eta) \over \zeta -\eta} = -{1 \over L} {1 \over \zeta-\eta} \langle \gamma | R_0(\zeta)-R_0(\eta)|\gamma \rangle \;,
\end{eqnarray}
where $\zeta, \eta$ is on the negative real axis and $L$ is the length of the curve $\Gamma$. Using the resolvent identity for the free resolvent, i.e., $R_0(\zeta)-R_0(\eta)= (\eta-\zeta) R_0(\zeta) R_0(\eta)$, we find
\begin{eqnarray}
\int_{\Delta} \int_{\Delta} d \zeta \; d \eta \; \bar{F}(\zeta) F(\eta) \left( {\Phi(\zeta)-\Phi(\eta) \over \zeta -\eta} \right) = {1 \over L} \left| \int_{\Delta} d \eta \; F(\eta) R_0(\eta) |\gamma \rangle \right|^2 > 0 \;,
\end{eqnarray}
where $R_0^{\dagger}(\eta)=R_0(\bar{\eta})=R_0(\eta)$. The positivity is preserved in the limiting case $\zeta \rightarrow \eta$ as well. This shows that the analytically continued function, say $\tilde{\Phi}$ is a Nevallina function. We denote the analytically continued function by the same letter $\Phi$ for simplicity. The aforementioned theorem can be generalized to the matrix valued function $\Phi_{ij}(E)$, as a result to ensure the holomorphicity we verify that:
\begin{eqnarray}
\int_{\Delta} \int_{\Delta} d \zeta \; d \eta \; \sum_{i,j=1}^{N} \bar{F}_i(\zeta) F_j(\eta) \left( {\Phi_{ij}(\zeta)-\Phi_{ij}(\eta) \over \zeta -\eta} \right) =  \left| \int_{\Delta} d \eta \; \sum_{i=1}^{N} {1 \over L_i} F_i(\eta) R_0(\eta) |\gamma_i \rangle \right|^2 > 0 \;,
\end{eqnarray}
and the principal matrix in all the other cases, including the relativistic extension of these problems, can be similarly analyzed. 
Hence, for a large region of the complex plane, which contains the negative real axis, the principal matrix is a matrix-valued holomorphic function so that its eigenvalues
and eigenprojections are holomorphic near the real axis \cite{Kato}. In fact, we get poles on the real axis for the eigenvalues and the residue calculus can be used to calculate the associated projections.

Let us consider the eigenvalue problem for the principal matrix depending on the real parameter $E$:
\begin{equation}
\Phi(E) A^{k}(E) = \omega^{k}(E) A^{k}(E) \;,
\end{equation} 
where $k=1,2, \ldots, N$ and we assume there is no degeneracy for simplicity (we consider the generic case). In order to simplify the notation, we sometimes suppress the variable $E$ in the equations, e.g., $A^{k}(E)=A^{k}$ and so on. Then, the bound state energies can be found from the zeroes of the eigenvalues $\omega$, that is,
\begin{equation}
\omega^k(E)=0 \;, \label{zeroesofeigenvalues}
\end{equation}
for each $k$. Thanks to Feynman-Hellmann theorem \cite{feynman, hellmann}, we have the following useful result  
\begin{equation}
{\partial \omega^{k} \over \partial E} = \langle A^{k}, {\partial \Phi \over \partial E} \; A^{k} \rangle \;,
\end{equation}
where $\langle ., . \rangle$ denotes the inner product on  $\mathbb{C}^N$. Using the expression of the principal matrices in all class of singular interactions described above and using the positivity of the heat kernel, it is possible to show that 
\begin{eqnarray}
{\partial \omega^{k} \over \partial E}<0 \;.
\end{eqnarray} 
This is an important result, since it implies that every eigenvalue cuts the real axis only once, that particular value  gives us a bound state if it  is below the spectrum of the free part. Moreover we deduce that the ground state  energy corresponds to the smallest eigenvalue of $\Phi$.

\section{Off-Diagonal Terms of the Principal Matrices in the Tunneling Regime}

\label{Off-Diagonal Terms of the Principal Matrix in the Tunneling Regime}

For simplicity, we assume that all binding
energies $E_{B}^{i}$'s  or/and $\lambda_i$'s are different. 
We consider the situation where the Dirac delta potentials (supported by points and curves) are separated far away from each other in the sense that the de Broglie wavelength of the particle is much smaller than the minimum distance $d$ between the point Dirac delta potentials or than the minimum distance between the delta potentials supported by non intersecting regular curves with finite length, namely 
\begin{eqnarray}
d \gg \lambda_{\mathrm{de\,Broglie}} \;, \label{tunneling regime1}
\end{eqnarray}
or in the semi-relativistic case, this can be stated as $d \gg \lambda_{\mathrm{Compton}}$. This regime can also be  defined in terms of the energy scales, namely
\begin{eqnarray}
{1 \over d^2} \ll E_{B} \;,
\end{eqnarray}
where $E_B$ is the minimum of the binding energies to the single delta potentials in the absence of all the others (recall that $\hbar=2m=1$). 

In the non-relativistic problem for point interactions in one and three dimensions, it is clear from the explicit form of the principal matrices (\ref{Phi 1D}), (\ref{principal matrix in 3D real}) all the off-diagonal terms are getting exponentially small as $d$ increases, i.e.,
\begin{equation}
|\Phi_{ij}(\nu)|= {\exp(-\nu d_{ij}) \over 2 \nu} \leq {\exp(-\nu d) \over 2\nu} \rightarrow 0 \;, 
\end{equation}
and
\begin{equation}
|\Phi_{ij}(\nu)|= {\exp(-\nu d_{ij}) \over 4 \pi d_{ij}} \leq {\exp(-\nu d) \over 4 \pi d} \rightarrow 0 \;,
\end{equation}
as $d \rightarrow \infty$. For point interactions in two dimensions, thanks to the upper bound of the Bessel function \cite{ErmanTurgut}, 
\begin{equation}
K_0(x) < {2 \over x} \exp(-x/2) \label{upperbound K_0} \;,
\end{equation}
for all $x$, the off-diagonal terms of the principal matrix (\ref{principal matrix in 2D real}) 
\begin{equation}
|\Phi_{ij}(\nu)| \leq {1 \over 2\pi} K_0(\nu|\mathbf{a_i} -\mathbf{a_{j}}|) \leq {1 \over 2\pi} K_0(\nu d) < {1 \over \nu \pi d} \exp(-\nu d) \;, 
\end{equation}
is going to zero exponentially as $d \rightarrow \infty$. In the above expressions for principal matrices, we have expressed them in terms of a real positive variable $\nu$ for simplicity. Not all the bound state spectra of the potentials we consider in this paper are negative, so it is not always useful to express the principal matrix in terms of a real positive  variable $\nu$. For that purpose, we will consider the principal matrices restricted to the real values, namely $z=E$, where $E$ is the real variable (not necessarily negative).

For point interactions in three dimensional hyperbolic manifolds,  the off-diagonal principal matrix restricted to the real values $E<\kappa^2$
\begin{eqnarray}
|\Phi_{ij}(E)| \leq \left( {\kappa \exp \left(- d  \sqrt{\kappa^2 -E}\right) \over 4 \pi \sinh \left(\kappa d \right) } \right)
\end{eqnarray}
is exponentially small as $d \rightarrow \infty$. Here $d$ is the minimum geodesic distance between the centers.

As for the point interactions in two dimensional hyperbolic manifolds, the off-diagonal principal matrix restricted to the real values $E<\kappa^2/4$
becomes 
\begin{eqnarray}
|\Phi_{ij}(E)| = {1 \over 2 \pi} \; Q_{ {1 \over 2} + \sqrt{-{E \over \kappa^2} + {1 \over 4}} } \left(\cosh(\kappa d(a_i,a_j) )\right) \;.
\end{eqnarray}
Using the series representation of the Legendre function of second kind \cite{Lebedev}
\begin{equation}
Q _ { v } ( \cosh \alpha ) = \sum _ { k = 0 } ^ { \infty } \frac { \Gamma ( k + v + 1 ) \Gamma \left( k + \frac { 1 } { 2 } \right) } { \Gamma \left( k + v + \frac { 3 } { 2 } \right) \Gamma ( k + 1 ) } e ^ { - \alpha ( 2 k + v + 1 ) } \;, \label{SeriesrepQ}
\end{equation}
where $v= {1 \over 2} + \sqrt{-{E \over \kappa^2} + {1 \over 4}} > 1 $ and $\alpha= \kappa d(a_i,a_j)$, and splitting the sum, we obtain
\begin{eqnarray}
\begin{aligned} & 
|\Phi_{ij}(E)| =    \frac { \Gamma ( v + 1 ) \Gamma \left( \frac { 1 } { 2 } \right) } { \Gamma \left(v + \frac { 3 } { 2 } \right) \Gamma ( 1 ) } e ^ { - \alpha (v + 1 ) } + \frac { \Gamma ( 1 + v + 1 ) \Gamma \left( 1 + \frac { 1 } { 2 } \right) } { \Gamma \left( 1 + v + \frac { 3 } { 2 } \right) \Gamma ( 1 + 1 ) } e ^ { - \alpha ( 2  + v + 1 ) } \\ & \hspace{2cm} + \sum _ { k = 2 } ^ { \infty } \frac { \Gamma ( k + v + 1 ) \Gamma \left( k + \frac { 1 } { 2 } \right) } { \Gamma \left( k + v + \frac { 3 } { 2 } \right) \Gamma ( k + 1 ) } e ^ { - \alpha ( 2 k + v + 1 ) } \;.
\end{aligned}
\end{eqnarray}
Since Gamma function is increasing on $[2,\infty)$, $\frac { \Gamma ( k + v + 1 ) \Gamma \left( k + \frac { 1 } { 2 } \right) } { \Gamma \left( k + v + \frac { 3 } { 2 } \right) \Gamma ( k + 1 ) } < 1 $ for all $k \geq 2$, and $v>1$, we can find an upper bound for the above the infinite sum as 
\begin{eqnarray}
e^{-4\kappa d- \kappa d(v+1)} \; \sum_{k=0}^{\infty} e^{-2k \kappa d} \;,
\end{eqnarray}
which is simply a geometric series. All these show that the off-diagonal principal matrix in two dimensional hyperbolic manifolds is exponentially small as $d \rightarrow \infty$ and the leading term is given by the first term of the series expansion. 

As for the delta interactions supported by curves, 
the minimum of the pairwise distances between the supports of Dirac delta potentials always exists since $d_{ij}(s,s')=\sqrt{|(\mathbf{\gamma_i(s)}-\mathbf{\gamma_{j}(s')})|}$ is a continuous function on compact interval $s \in [0,L]$,
so we have
\begin{equation}
|(\mathbf{\gamma_i}(s_i)-\mathbf{\gamma_{j}}(s'_{j}))|^2 \geq \min_{s_i,s'_{j}} |(\mathbf{\gamma_i}(s_i)-\mathbf{\gamma_{j}}(s'_{j}))|^2 := d_{ij} \geq \min_{ij} d_{ij} := d \;,
\end{equation}
for $i \neq j$. Then, 
\begin{equation}
|\Phi_{ij}(E)| \leq \sqrt{L_i \, L_{j}} \int_{0}^{\infty} dt \; {e^{-d^2/4t + t E} \over 4 \pi t}= {\sqrt{L_i \, L_{j}} \over 2 \pi} K_0(\sqrt{-E} d) \;.
\end{equation}
Due to the upper bound of the Bessel function (\ref{upperbound K_0}), the off-diagonal principal matrix is going to zero as $d \rightarrow \infty$.

Similarly, the explicit forms of the off-diagonal parts of the principal matrices (\ref{relativistic Phi}) and (\ref{Phi relativistic point 2}) in the relativistic cases are exponentially going to zero as $d \rightarrow \infty$ (by assuming the order of the limit and the integral can be interchanged). For the other relativistic cases (including the relativistic delta potentials supported by curves), the off diagonal terms of the principal matrices can also be shown to be exponentially small.

Therefore, we see that the principal matrices for all the above models are diagonally dominant in the ``large" separation regime. However, the exponentially small off-diagonal terms are not analytic in the small parameter (${1 \over E_{B} d^2}$). Nevertheless, we can keep track of  small values of the off-diagonal terms by introducing an artificial parameter $\epsilon$ in order to control the orders of terms in the perturbative expansion, that we are going to develop in the next section.

\section{Splitting in Bound State Energies through Perturbation Theory}
\label{Splitting in Bound State Energies through Perturbation Theory}

Let us consider the family of principal matrices restricted to the real axis $E$:
\begin{equation}
\Phi(E)= \Phi_{0}(E) + \epsilon \, \delta\Phi(E) \;,
\end{equation}
where $\Phi_{0}$ is the diagonal part of the principal matrix, and $\delta\Phi$ is off-diagonal part of it and this is the ``small" correction (perturbation) to the diagonal part. Since $\Phi(E)$ is symmetric (Hermitian), we can apply standard
perturbation techniques to the principal matrix \cite{Kato, Galindo Pascual, reedsimonv4}. 
For this purpose, let us assume we can expand the eigenvalues and eigenvectors as follows:  
\begin{eqnarray}
\omega^{k} & = & \omega^{k}_0 + \epsilon \; \omega^{k}_{1} + \epsilon^2 \;\omega^{k}_{2} + \ldots \cr  A^{k} & = & A^{k}_{0} + \epsilon \; A^{k}_{1} +  \epsilon^2 \; A^{k}_{2} + \ldots \;,
\end{eqnarray}
for each $k$.  
The solution to the related unperturbed eigenvalue problem
\be  \Phi_{0}
A^{k}_{0}= \omega^{k}_{0}
A^{k}_{0} \;, \ee
is given by
\be \omega^{k}_{0} = [\Phi_0]_{kk} \;. \ee
Once we have found the eigenvalues and eigenvectors of
the diagonal part of the principal matrix or unperturbed
eigenvalue problem, we can perturbatively solve the full problem.
The standard perturbation theory gives us the eigenvalues $\omega^k$ up to second order:
\beqs \omega^{k}_{1} (E) &=&  \langle A^{k}_{0}(E), \delta \Phi(E) A^{k}_{0}(E) \rangle= \left[\delta
\Phi(E)\right]_{kk}\;, \label{perturbation eigenvalues1} \eeqs
\beqs
\omega^{k}_{2}(E) &=&
\sum_{\substack{l=1 \\
l \neq k}}^N { \left| \langle A^{k}_{0}(E), \delta \Phi(E) A^{k}_{0}(E)\rangle \right|^2 \over
\omega^{k}_{0}(E)-\omega^{l}_{0}(E)} \cr &=& \sum_{\substack{l=1 \\
l \neq k}}^N {\Phi_{lk}(E) \Phi_{k l}(E) \over
\omega^{k}_{0}(E)-\omega^{l}_{0}(E) } \;.
\label{perturbation eigenvalues} \eeqs
and the first order correction to the eigenvectors $A^k$ is given by 
\begin{eqnarray}
A^{k}_{1}(E)= \sum_{\substack{j=1 \\ j \neq k}}^N {\delta \Phi_{jk}(E) \over \omega_{0}^{k}(E)-\omega_{0}^{j}(E)} \;A_{0}^{j}(E) \;.\label{1storderA}
\end{eqnarray}

Since the bound state energies are determined from the solution of equation (\ref{zeroesofeigenvalues}), the bound state energies in the zeroth order approximation can easily be found from $\omega^{k}_{0}(E)=0$. The solution is given by
\be E= E^{k}_{0}=E_{B}^{k} \;, \ee
and the corresponding eigenvector is
\begin{equation} A^{k}_{0}(E_{B}^{k}) \equiv A^{k}_{0} \equiv
\mathbf{e}^{k} \equiv
\left(%
\begin{array}{cc}
  0 \cr \vdots \cr 1 \cr \vdots \cr 0
\end{array}%
\right)\;,
\end{equation}
where $1$ is located in the $k$th position of the column and other
elements of it are zero or we can write 
\begin{eqnarray}
    A^{k i}_{0} = e^{k}_i=
\delta_{k i} \;. \label{uniteigenvector}
\end{eqnarray}
Here $e^{k}_{i}$ s form a complete orthonormal set
of basis.
\be \sum_{i=1}^{N} e^{k}_{i}e^{l}_{i}= \delta_{kl} \;. \ee
The bound state energies to the full problem up to the second order
is then determined by solving the following equation
\be \omega^{k}(E) = \omega^{k}_{0}(E)  + \epsilon^2 \, \omega^{k}_{2}(E) = 0
\;, \label{perturb eigenvalue eq}\ee
where we have used the first order result
\begin{equation}
    \omega^{k}_{1}=0
\end{equation} 
from the Equation (\ref{perturbation eigenvalues1}).

Let us now expand $\omega^{k}_{0}(E)$ and $\Phi_{kl}(E)$ for $k
\neq l$ around $E=E_{B}^k$:
\beqs  \omega^{k}_{0}(E) &=&  \left. {\partial
\omega^{k}_{0}(E) \over
\partial E} \right|_{E=E_{B}^k} \delta E^k + \mathcal{O}((\delta E^k)^2) \;, \cr
\Phi_{kl}(E) &=& \Phi_{kl}(E_{B}^k) + \left.
{\partial \Phi_{kl}(E) \over \partial E}
\right|_{E=E_{B}^k} \delta E^k + \mathcal{O}((\delta E^k)^2)
\;, \label{perturb taylor exp} \eeqs
where $\omega^{k}_{0}(E_{B}^k)=0$. If we
substitute (\ref{perturb taylor exp}) into (\ref{perturb
eigenvalue eq}) and (\ref{perturbation eigenvalues}), and use
Feynman-Hellman theorem given in previous section, the condition (\ref{perturb eigenvalue eq}) up to the second order turns out be
\begin{eqnarray}
\begin{aligned} 
& \left. {\partial \Phi_{kk}(E)
\over
\partial
E} \right|_{E=E_{B}^{k}} \delta E^k - \epsilon^2 \; \sum_{\substack{l=1 \\
l \neq k}}^N {1 \over \Phi_{ll}(E_{B}^{k})}
\Bigg[\Phi_{kl}(E_{B}^k) \Phi_{lk}(E_{B}^k)  \\ & \hspace{3cm} + \Bigg(
\Phi_{kl}(E_{B}^k) \left. {\partial \Phi_{lk}(E) \over
\partial E} \right|_{E=E_{B}^k} + \Phi_{lk}(E_{B}^k) \left. {\partial \Phi_{kl}(E) \over
\partial E} \right|_{E=E_{B}^k}\Bigg) \delta E^k \Bigg]  \\ & \hspace{2cm} \times \Bigg[1 +
{1 \over \Phi_{ll}(E_{B}^k)} \Bigg( \left. {\partial
\Phi_{ll}(E) \over
\partial E} \right|_{E=E_{B}^k}- \left. {\partial \Phi_{kk}(E) \over
\partial E} \right|_{E=E_{B}^k} \Bigg) \delta E^k \Bigg]^{-1}  + \mathcal{O}((\delta E^k)^2) = 0 \;. 
\end{aligned}
\end{eqnarray}
If we also expand the last factor in the powers of $(\delta E^k)$
and ignore the second order terms and combine the terms using the
symmetry property of principal matrix, we find
\beqs 
\begin{aligned}
& \Bigg[ \left. {\partial \Phi_{kk}(E) \over
\partial E} \right|_{E=E_{B}^k} + \epsilon^2 \; \sum_{\substack{l=1 \\
l \neq k}}^N  {\Phi_{kl}(E_{B}^k) \Phi_{lk}(E_{B}^k) \over
\Phi_{ll}^2(E_{B}^k)} \Bigg( \left. {\partial \Phi_{ll}(E)
\over
\partial E} \right|_{E=E_{B}^k} \\ & \hspace{5cm} - \left. {\partial \Phi_{kk}(E) \over
\partial E} \right|_{E=E_{B}^k} \Bigg) - 2 \sum_{\substack{l=1 \\
l \neq k}}^N  {\Phi_{kl}(E_{B}^k) \over \Phi_{ll}(E_{B}^k)}
\left. {\partial \Phi_{lk}(E) \over
\partial E} \right|_{E=E_{B}^k} \Bigg] \delta E^k \\ & \hspace{3cm} = \epsilon^2 \; \sum_{\substack{l=1 \\
l \neq k}}^N  {\Phi_{kl}(E_{B}^k) \Phi_{lk}(E_{B}^k) \over
\Phi_{ll}(E_{B}^k)} + \mathcal{O}((\delta E^k)^2)\;. 
\end{aligned}
\eeqs
Ignoring the second and third terms on the left hand side of the
equality (this is guaranteed by the assumption $\Phi_{kk}(E_{B}^k) \gg
|\Phi_{kl}(E_{B}^k)|$) and setting $\epsilon=1$, we get the change in $E^k$ (to first order) as,
\be \delta E^k \simeq   \left(\left. {\partial
\Phi_{kk}(E) \over
\partial E} \right|_{E=E_{B}^k} \right)^{-1} \sum_{\substack{l=1 \\
l \neq k}}^N  {\Phi_{kl}(E_{B}^k) \Phi_{lk}(E_{B}^k) \over
\Phi_{ll}(E_{B}^k)} + \mathcal{O}((\delta E^k)^2)\;.
\label{perturbation result} \ee
This is our main formula for all types of singular interactions we consider. It is striking  that it contains the information about the tunneling regime.

\section{Explicit Examples for the Splitting in the Energy}
\label{Explicit Examples for the Splitting in the Energy}

Let us now compute explicitly how the bound state energies change in the
tunneling regime for the above class of singular potentials.

For point Dirac delta potentials in one dimension, the bound state energies are negative so $E_{B}^k= -|E_{B}^k|$ and
\begin{equation}
\delta E^k \simeq \sqrt{|E_{B}^k|} \sum_{\substack{l=1 \\
l \neq k}}^N {1 \over \left( {1 \over \lambda_l} - {1 \over 2 \sqrt{|E_{B}^k|}}\right)} \; \exp \left(-2 \sqrt{|E_{B}^k|} \; |a_k-a_l|\right)\;,
\end{equation}
in the tunneling regime $ d \sqrt{|E_{B}|} \gg 1$.

For point Dirac delta potentials in two dimensions, the bound state energies are negative and 
\begin{equation}
\delta E^k \simeq \sum_{\substack{l=1 \\
l \neq k}}^N {2 \pi \over \sqrt{|E_{B}^k|} |\mathbf{a_k}-\mathbf{a_l}| \log(E_{B}^k / E_{B}^l)} \; \exp\left( -2 \sqrt{|E_{B}^k|} |\mathbf{a_k}-\mathbf{a_l}| \right) \;,
\end{equation}
again in the tunneling regime. Here we have used the asymptotic expansion of the modified Bessel function of the third kind $K_0(x) \approx \sqrt{{\pi \over 2 x}} \; \exp(-x) $ for $x \gg 1$ \cite{Lebedev}.

In three dimensions, we have
\begin{equation}
\delta E^k \simeq \sum_{\substack{l=1 \\
l \neq k}}^N {2 \sqrt{|E_{B}^k|} \over 4 \pi^2 |\mathbf{a_k}-\mathbf{a_l}|^2} \; {\exp\left(-2 \sqrt{|E_{B}^k|} |\mathbf{a_k}-\mathbf{a_l}| \right) \over \left( \sqrt{|E_{B}^k|}- \sqrt{|E_{B}^l|} \right)} \;.
\end{equation}
For point interactions in three dimensional hyperbolic manifolds, the bound state energies are below $\kappa^2$ (see \cite{ErmannumberH} for details) and
\begin{equation}
\delta E^k \simeq \sqrt{\kappa^2 -E_{B}^k} \; \sum_{\substack{l=1 \\
l \neq k}}^N {4 \kappa^2 \over \sqrt{\kappa^2-E_{B}^k} - \sqrt{\kappa^2-E_{B}^l} } \; \exp \left( -2 d(a_k,a_l) \left( \kappa + \sqrt{\kappa^2 -E_{B}^k} \right) \right) \;,
\end{equation}
in the tunneling regime. Here we have used $\sinh^2 x \approx {e^{2x} \over 4} $ as $x \gg 1$. 

For point interactions in two dimensional hyperbolic manifolds, the bound state energies are below $\kappa^2/4$ (see \cite{ErmannumberH}) and    
\begin{eqnarray}
\begin{aligned} & 
\delta E^k \simeq {2 \kappa^2 \sqrt{{1 \over 4}- {E_{B}^k \over \kappa^2}} \over \psi^{(1)} \left({1 \over 2} + \sqrt{{1 \over 4}- {E_{B}^k \over \kappa^2}}\right)} \sum_{\substack{l=1 \\
l \neq k}}^N {1 \over \psi \left({1 \over 2} + \sqrt{{1 \over 4}- {E_{B}^k \over \kappa^2}}\right) - \psi \left({1 \over 2} + \sqrt{{1 \over 4}- {E_{B}^l \over \kappa^2}}\right) } \\ & \hspace{1cm}\times \sum_{m=0}^{\infty} {\Gamma \left(m + {3 \over 2} +\sqrt{{1 \over 4}- {E_{B}^k \over \kappa^2}} \right) \Gamma(m+{1 \over 2}) \over \Gamma \left(m + 2 +\sqrt{{1 \over 4}- {E_{B}^k \over \kappa^2}} \right) \Gamma(m+1)} \; \exp \left(- \kappa d(a_k,a_l) \left( 2m+{3 \over 2} + \sqrt{{1 \over 4}- {E_{B}^k \over \kappa^2}} \right) \right) \;,
\end{aligned} 
\end{eqnarray}
where $\psi^{(1)}$ is the polygamma function and we have used the infinite series representation of the Legendre function of second kind (\ref{SeriesrepQ}).

For semi-relativistic point interactions in one dimensions, the bound state energies are below $m$. Let us first find explicitly integrals in the off-diagonal part of the principal matrix asymptotically 
\begin{equation}
\frac { 1 } { \pi } \int _ { m } ^ { \infty } d \mu e ^ { - \mu \left| a _ { k } - a _ { l } \right| } \frac { \sqrt { \mu ^ { 2 } - m ^ { 2 } } } { \mu ^ { 2 } - m ^ { 2 } + (E_{B}^k)^ { 2 } } 
\end{equation}
in the tunneling regime $m d \gg 1$. For this purpose, let us rescale the integration variable $s=\mu/m$ so that the above integral becomes ${m^2 \over \pi} \int_{1}^{\infty} {e^{-s m |a_k-a_l|} \sqrt{s^2-1} \over m^2(s^2-1)+ (E_{B}^{k})^2}$.  
Note that $-s$ in the exponent has its maximum at $s=1$ on the interval $(1, \infty)$. Then, only the vicinity of $s=1$ contributes to the full asymptotic expansion of the integral for large $m|a_k-a_l|$. Thus, we may approximate the above integral by $ {m^2 \over \pi} \int_{1}^{\epsilon} {e^{-s m |a_k-a_l|} \sqrt{s^2-1} \over m^2(s^2-1)+ (E_{B}^{k})^2}$, where $\epsilon>1$ and replace the function ${\sqrt{s^2-1} \over m^2(s^2-1)+ (E_{B}^{k})^2}$ in the integrand by its Taylor expansion \cite{bender}. It is important to emphasize that the full asymptotic expansion of this integral as $m|a_k-a_l| \rightarrow \infty$ does not depend on $\epsilon$ since all other integrations are subdominant compared to the original integral. Hence, we find
\beqs
{m^2 \over \pi} \int_{1}^{\epsilon} {e^{-s m |a_k-a_l|} \sqrt{s^2-1} \over m^2(s^2-1)+ (E_{B}^{k})^2} & \sim & {m^2 \over \pi} \int_{1}^{\epsilon} d s \; e^{- s m |a_k-a_l|} \; {\sqrt{2} \; \sqrt{s-1} \over  (E_{B}^{k})^2} 
\cr & \sim & 
{m^2 \over \pi} \int_{1}^{\infty} d s \; e^{- s m |a_k-a_l|} \; {\sqrt{2} \; \sqrt{s-1} \over  (E_{B}^{k})^2}
\cr &\sim & {1 \over \sqrt{2\pi}} \left( {m \over E_{B}^k}\right)^2 {1 \over m |a_k-a_l|^{3/2}} \; \exp\left( -m |a_k-a_l| \right)  \;,
\eeqs
where we have used the fact that the contribution to the integral outside of the interval $(1,\epsilon)$ is exponentially small. Substituting this result into Eq. (\ref{perturbation result}), we find 
\begin{eqnarray}
\delta E^k \simeq \left( \varphi'(E_{B}^k)\right)^{-1} \; \sum_{\substack{l=1 \\
l \neq k}}^N {1 \over 2\pi} \left( {m \over E_{B}^k}\right)^4 {1 \over m |a_k-a_l|^{3}} \;{1 \over \varphi(E_{B}^k)-\varphi(E_{B}^l)} \exp\left( -2 m |a_k-a_l| \right) 
\end{eqnarray}
when $E_{B}^k <0$ and
\begin{eqnarray}
\begin{aligned} & 
\delta E^k \simeq \left( \varphi'(E_{B}^k)\right)^{-1} \; \sum_{\substack{l=1 \\
l \neq k}}^N {1 \over \varphi(E_{B}^k)-\varphi(E_{B}^l)}  \bigg( {e^{-\sqrt{m^2 -(E_{B}^{k})^2}|a_k-a_l|} E_{B}^k \over \sqrt{m^2-(E_{B}^{k})^2}} \\ & \hspace{4cm}+ {1 \over \sqrt{2\pi}} \left({m \over E_{B}^k}\right)^2 {1 \over m |a_k-a_l|^{3/2}} \; \exp\left( - m |a_k-a_l| \right)  \bigg)^2
\end{aligned}
\end{eqnarray}
when $E_{B}^k>0$.

For the field theory motivated relativistic version we can use a saddle point approximation, assuming that tunneling condition, given by 
$\sqrt{m^2-(E^k_B)^2} d_{ij}>>1$ is satisfied. Here it is enough to consider the function $m(1+s^2)^{1/2} -E^k_B s$ and expand it around the maximum 
$E^k_B/\sqrt{m^2- (E^k_B)^2}$. The denominator can be replaced by its value at the maximum, we find that the leading behaviour goes as 
\begin{equation}
   \Phi_{ij}(E_B^k)\sim -{1\over 2\pi}{\sqrt{m^2- (E^k_B)^2}\over m}e^{-d_{ij}\sqrt{m^2-(E_B^k)^2}}  \int_{-E^k_B/\sqrt{m^2- (E^k_B)^2}}^\infty d\xi\,  e^{-d_{ij}[m^2- (E^k_B)^2]^{3/2}{\xi^2\over 2m^2}},
\end{equation}
(assuming that $E_B^k d_{ij}$'s remain large) evaluating the integral we end up with,
\begin{equation}
    \Phi_{ij}(E^k_B)\sim -{1\over\sqrt{ 2\pi} }{1 \over [ d_{ij}\sqrt{m^2- (E^k_B)^2}]^{1/2}} e^{-d_{ij}\sqrt{m^2-(E_B^k)^2}}.
\end{equation}
Once we obtain the off-diagonal terms responsible for the tunneling contributions, calculating the derivatives of the diagonal parts are simple,
\begin{equation}
    {\partial \Phi_{ii} (E)\over \partial E}\Big|_{E=E_B^k}= -{1\over 2\pi} {1\over m-E_B^k}.
\end{equation}
Substituting these expressions into the general formulae we have derived,  gives the tunneling contribution  to energy levels that leads to small shifts in the binding energies.

For Dirac delta potentials supported by curves in two dimensions:
we define a kind of center of mass by 
\begin{equation}
{\mathbf x}_i= {1\over L_i }\int_{\Gamma_i} ds_i \; {\mathbf\gamma}(s_i) \;,
\end{equation}
and write
\begin{equation}
 |{\mathbf \gamma} (s_i) -{\mathbf \gamma_j}(s_j)|=
|{\mathbf \gamma} (s_i) -{\mathbf x}_i -{\mathbf \gamma_j}(s_j)+{\mathbf x}_j +({\mathbf x}_i -{\mathbf x}_j)| \;,
\end{equation}
in the argument of the functions in the principal matrix.
When we evaluate the expressions we expand these terms by keeping only first order terms in the small quantities. The resulting 
Bessel functions can be expanded again to find the leading corrections for the curve to curve interaction terms.
We use the expression above for the off diagonal terms and define ${d_{ij}}=|{\bf x}_i-{\bf x}_j|$ for simplicity and introduce a unit vector as $\hat {\bf d_{ij}}$ in a similar way. As a result we have the leading order expansion,
\begin{equation}
K_0(\sqrt{-E} d_{ij})-K_1(\sqrt{-E} d_{ij}){1\over d_{ij} }  \Big[ \hat {\bf d_{ij}} \cdot ( \gamma_i(s_i)-{\bf x}_i) - \hat {\bf d_{ij}} \cdot  (\gamma_j(s_j)-{\bf x}_j) \Big] \;.
\end{equation}
When we insert this into $\Phi_{ij}$
expression and integrate over the curve, 
we find
\begin{equation}
\int ds_i \; \hat {\bf d_{ij}} \cdot ( \gamma_i(s_i)-{\bf x}_i)=
\hat {\bf d_{ij}} \cdot \int ds_i \; (\gamma_i(s_i)-{\bf x}_i)=0 \;,
\end{equation}
and similarly for the other part.
Thus we see that the only contribution comes from the second order which we neglect for our purposes. However a systematic expansion in powers of ${1\over d_{ij}}$
can be developed for higher order correction as described.
Using the asymptotic expansion of $K_0(z)$ for large values of $z$ \cite{Lebedev}, 
\begin{equation}
    K_\nu(z) \sim \sqrt{\pi \over 2 z} e^{-z}, \label{largeasymptoticbessel}
\end{equation}
for all $\nu \geq 0$ we get from (\ref{perturbation result}) a more elegant expression,
\begin{equation}
\delta E^k \simeq  \left( \left. {\partial
\Phi_{kk}(E) \over
\partial E} \right|_{E=E_{B}^k} \right)^{-1} \sum_{\substack{l=1 \\
l \neq k}}^N  { \left(L_k L_l/8 \pi \sqrt{|E_{B}^k|}d_{kl}\right)  \over
\left(\Phi_{ll}(E_{B}^k)\right)} \; \exp\left(-2\sqrt{|E_{B}^k|}d_{kl} \right) + \mathcal{O}((\delta E^k)^2) \;,
\end{equation}
where $\Phi_{ll}$ and its derivative at $E_{B}^k$ can be computed from the explicit expression of the principal matrix (\ref{Phi_curve2D}).
For Dirac delta potentials supported by curves in three dimensions, there is really no change, since renormalization is required only for  the diagonal parts, we have the off-diagonal expressions already in a simpler form, as a result of the above analysis, the leading order expression is found to be,
\begin{eqnarray}
\delta E^k \simeq  
 \left( \left. {\partial
\Phi_{kk}(E) \over
\partial E} \right|_{E=E_{B}^k} \right)^{-1} \sum_{\substack{l=1 \\
l \neq k}}^N  {\left(L_k L_l/16\pi^2 d_{kl}^2 \right)  \over
\left(\Phi_{ll}(E_{B}^k)\right)} \; \exp\left(-2\sqrt{|E_{B}^k|}d_{kl} \right) + \mathcal{O}((\delta E^k)^2) \;,
\end{eqnarray}
where $\Phi_{ll}$ and its derivative at $E_{B}^k$ can be computed from the explicit expression of the principal matrix (\ref{principal_curve_3D}).

In a similar way, we look at the tunneling correction to bound state energies for relativistic particle coupled to Dirac potentials supported over curves.
Again we use the  approximation that the separation of the curves are large and the extend of the curves compared to these distances are small. This is not the only possible approximation, one can envisage a situation in which the separations are large but the extend of the curves are also large. The essential ideas are captured by our example so to achieve technical simplicity we keep this approximation. 
Essential point is to expand the off-diagonal terms in the leading order.
By scaling $t$ variable in the integral we can write $\Phi_{ij}(E_B^{k})$
term as,
\begin{eqnarray}
\Phi_{ij}(E_B^{k}) &=&-{m\over \sqrt{2L_iL_j} \pi^2 }\int_0^\infty dt \int_{\Gamma_i\times \Gamma_j}ds_i ds_j \;  {K_1\big(m|\gamma(s_i)-\gamma(s_j)|\sqrt{t^2 +1}\big)\over \sqrt{t^2+1 }} e^{E_B^{k} t|\gamma(s_i)-\gamma(s_j)|}\nonumber\\
&\sim& -\int_{\Gamma_i\times \Gamma_j}ds_i ds_j \; {m^{1/2}\over 2\sqrt{|\gamma(s_i)-\gamma(s_j)|} (L_iL_j)^{1/2} \pi^{3/2} } \int_0^\infty dt \; {e^{-|\gamma(s_i)-\gamma(s_j)|[m\sqrt{t^2+1}-E_B^{k} t]}\over (t^2+1)^{3/4}},
\end{eqnarray}
where in the second line we used the asymptotics of $K_1$
for large argument (\ref{largeasymptoticbessel}). 
We may now use the same argument by means of the center of mass of the curves to define center to center distances and  expand around the center of mass, not surprisingly we again find that the first order corrections become zero, only the center to center distance matters.
Therefore, to leading order we have  a simpler expression,
\begin{equation}
\Phi_{ij}(E_B^{k})\sim -{m^{1/2} (L_iL_j)^{1/2}\over 2 \pi^{3/2} d_{ij}^{1/2} } \int_0^\infty dt  \; {e^{-d_{ij}[m\sqrt{t^2+1}-E_B^{k} t]}\over (t^2+1)^{3/4}}  
.\end{equation}
This is of the type we have  worked out for the semi-relativistic particle, and in the same manner, a saddle point approximation can be applied  in a simple way, resulting
\begin{equation}
    \Phi_{ij}(E_B^{k})\sim -{(L_iL_j)^{1/2}\over \sqrt{2} \pi d_{ij}} \; e^{-d_{ij}\sqrt{m^2-(E_B^{k})^2}} \;.
\end{equation}
We may now employ our general expressions to find the tunneling corrections. The derivative of the diagonal term can be simplified by means of ${\partial K_0(z)\over \partial z}=-K_1(z)$.

\section{Degenerate Case and Wave Functions for Point Interactions}
\label{Degenerate Case and Wave Functions for Point Interactions}

\begin{figure}[ht]
\centering
\includegraphics[scale=0.6]{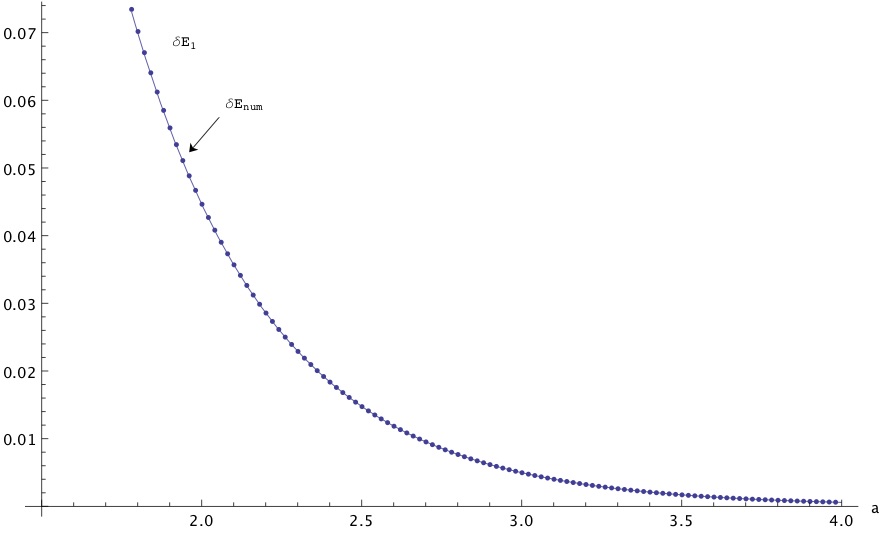}
\caption{Numerical and First order Perturbation Results for the Splitting in the Energy as a function of $a$ for $\mu=1$ unit in two dimensions.}
\label{Figdelta2D}
\end{figure}

Let us now compute the energy splitting of two equal strength delta functions supported by the points $-\mathbf{a}$ and $\mathbf{a}$ in two dimensions.
This is very similar to the double well problem we discuss in the introduction, yet this version can be solved exactly. The approximation we use  corresponds to  the standard WKB approach.
Let us recall that when we have two degenerate eigenvalues
\begin{equation}
    \omega^1_0(E)=\omega^2_0(E), 
\end{equation}
the degeneracy is lifted by the diagonal perturbation and as is well known,
diagonalizing the perturbation matrix in the degeneracy subspace gives us the first order correction:
\begin{eqnarray}
    \omega^1_1(E) & = & + |\Phi_{12}(E)|\;, \cr  \omega^2_1(E) & =& - |\Phi_{12}(E)| \;.
\end{eqnarray}
If we call the common bound state as  $E_B$, for $k=1,2$ to get the first order correction we truncate the eigenvalue equations as, 
\begin{equation}
    \omega^k_0(E_B+ E^k_1)+\omega^k_1(E_B)=0
\end{equation}
which leads to 
\begin{equation}
    E^k_1\sim (-1)^{k+1} 2 |E_B| K_0(2\sqrt{|E_B|}a)\sim (-1)^{k+1} {|E_B|^{3/4} \sqrt{\pi} \over \sqrt{a} } \; e^{-2\sqrt{|E_B|}a} \;,
\end{equation}
where we have used the asymptotic expansion of $K_0$ given by (\ref{largeasymptoticbessel}). Thus the splitting is given by 
\begin{equation}
    \delta E_1 = E_{1}^{1}-E_{1}^{2} \sim 2 {|E_B|^{3/4} \sqrt{\pi} \over \sqrt{a} } \; e^{-2\sqrt{|E_B|}a}  \;,
\end{equation}
which should be compared with the usual one-dimensional double well potential splitting given in the introduction.
Note that in the former case, the strength of each harmonic well is proportional to the square of the separation therefore the initial energy level is not independent as in the delta function case and is proportional to the square of the separation. 
the exponent thus gets the square of the distance as the suppression factor, if we assume that $E_B\sim |a|^2$ one can see that the exponents behave in a similar way. Actually, one can also compare the first order perturbation result for the splitting $\delta E_1$ with the numerical result by solving $\det \Phi(\nu)= \ln(\nu/\mu)-\pm K_0(2a \nu)=0$ numerically for each $a$ by Mathematica (see Figure \ref{Figdelta2D}). We assume that $a>e^{\gamma}$ in order to guarantee the existence of the second bound states, where $\gamma$ is the Euler's constant.

\begin{figure}[ht]
\centering
\includegraphics[scale=0.6]{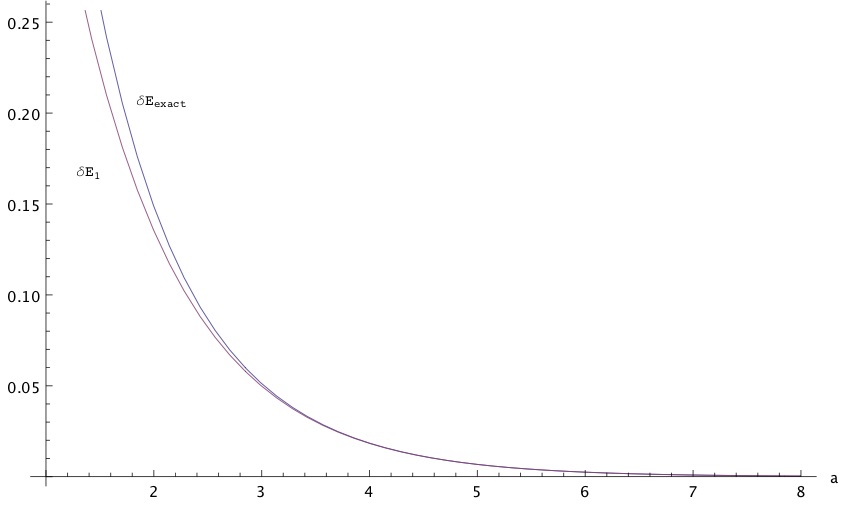}
\caption{Exact and First order Perturbation Results for the Splitting in the Energy as a function of $a$ for $\lambda=1$ unit in one dimension.}
\label{Figdelta1D}
\end{figure}

The same method can also be applied to the one-dimensional case. In the symmetrically placed Dirac delta potentials with equal strengths $\lambda$, the exact bound state energies when they are sufficiently far away from each other (when $a>1/\lambda$, there are two bound state energies) can analytically be computed \cite{EGU epp}
\begin{eqnarray}
    E_{\pm} = -\left({\lambda \over 2} + {1 \over 2a} W\left(\pm a \lambda e^{-a \lambda}\right) \right)^2 \;, \label{exact energ 1D}
\end{eqnarray}
where $W$ is the Lambert $W$ function \cite{LambertW}, which is defined as the solution $y(x)$ of the transcendental equation $y e^y = x$. From (\ref{principal matrix 1D z}), the principal matrix in this case reads
\begin{equation}
\Phi_{i j}(E) = \left\{ \begin{array} { l l } { {1 \over \lambda} -  {1 \over 2 \sqrt{-E}} } & { \text { if } i = j } \\ { -{1 \over 2 \sqrt{-E}} e^{-2a \sqrt{-E}}} & { \text { if } i \neq j } \end{array} \right. \;. \label{prncipal matrix 1d}
\end{equation}
Then, the first order perturbation result following the above procedure gives
\begin{eqnarray}
    \delta E_1= \lambda^2 e^{-a \lambda} \;,
\end{eqnarray}
where we have used well-known result $E_B =-{\lambda^2 \over 4}$. Then, one can easily find the error between the exact result $\delta E_{exact}=E_{+}-E_{-}$ and the first order perturbation result $\delta E_1$ in the splitting of the energy, see the Figure \ref{Figdelta1D}.

The three dimensional case can also be studied in this way and we can similarly solve $\det \Phi(\nu)=(\nu-\mu)-\pm {1 \over 2a} e^{-2a \nu}$ in terms of the Lambert W function and compare with the first order order perturbation result for the splitting in the energy (Figure \ref{Figdelta3D}):

\begin{figure}[ht]
\centering
\includegraphics[scale=0.6]{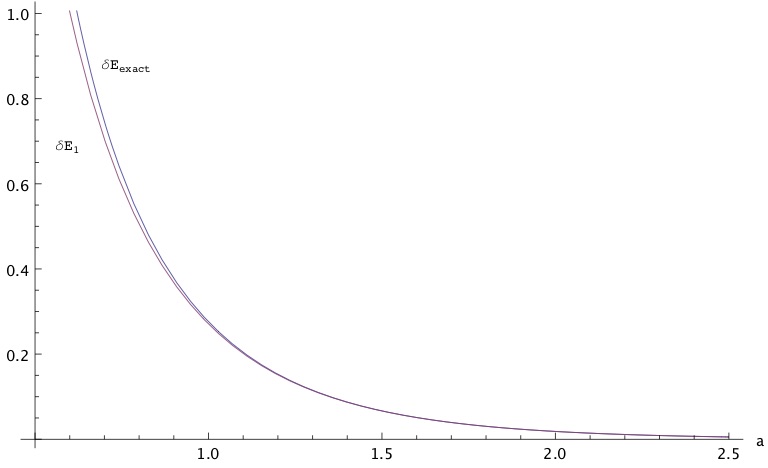}
\caption{Exact and First order Perturbation Results for the Splitting in the Energy as a function of $a$ for $\mu=1$ unit in three dimensions.}
\label{Figdelta3D}
\end{figure}
Here we assume that $a>1/2\mu$ in order to guarantee the existence of second bound states.

Let us emphasize that in the usual WKB approach one  constructs the wave functions in classically allowed and forbidden regions respectively and use a subtle argument to connect the different regions. In this case, there is really no forbidden region, except the supports of the attractive regions. Indeed right there classically there is no sensible way to define the motion of a particle. Nevertheless, it is possible to find the effect of tunneling for the wave functions from our formalism. 
It relies on the first order corrections to the eigenstates of the principal operator, 
notice that an expansion of the eigenstates of the principal operator can be found in the non-degenerate case as
\begin{equation}
    A^k(E_{B}^k)  =A^k_0(E_{B}^k) + \sum_{r \neq k} {\langle A^k_0(E_{B}^k), \delta \Phi_{kr}(E_{B}^k) A^r_0(E_{B}^k)\rangle \over \omega^k_0(E_{B}^k)-\omega^r_0(E_{B}^k)} A_0^r(E_{B}^k) \;.
\end{equation}
Note that  to this order normalization is not important, moreover we do not need to use a subtle argument about the shift of the eigenvalues since the change of eigenvalue is already second order in the exponentially small quantities, any such correction will be of lower order as we have seen in the shift of energy calculations.

It is well-known that the wave function of the system associated with the bound states can be found from the explicit expression of the resolvent formula. Since the eigenvalues
are isolated we can find the projections onto the subspace corresponding to this eigenvalue by the following contour integral (Riesz Integral representation) \cite{reedsimonv4}:
\be \mathbb{P}_{k} = - {1\over 2\pi i} \oint_{C_k}
\mathrm{d} z \;  R(z), \ee
where $C_k$ is a small contour enclosing the  isolated
eigenvalue, say $E_k$. We note that the free resolvent  does not contain any poles on the negative real axis for the Dirac delta potentials supported by points,
so all the poles on the negative real axis will come from the
poles of inverse principal matrix $\Phi^{-1}(z)$. 
Since the principal matrix is self-adjoint on the real axis, we can apply the spectral theorem. Moreover, its eigenvalues and eigenprojections are holomorphic near the real axis, as emphasized in section \ref{Bound State Spectrum}. Then, we can write the spectral resolution of the inverse principal matrix,
\be
 \Phi^{-1}_{ij}(z)=\sum_{k} {1 \over \omega^k(z)} \mathbb{P}_k(z)_{ij}
\;, \ee
where $\mathbb{P}_k(z)_{ij}=\overline{A^{i k}(z)} A^{j k}(z)$, $A^{k i}(z)$ is the
normalized eigenvector corresponding to the eigenvalue
$\omega^k(z)$. Then, from the residue theorem, we find the square integrable wave function associated with the bound state energy $E_k$ as 
\be
\begin{aligned}
\psi_k(\mathbf{x}) = & \alpha \; \sum_{i=1}^N R_0(\mathbf{x}, \mathbf{a_i};E_k) \, A^{k i}(E_k)  \;,
\label{wavefunction heat delta} \end{aligned} \ee
where $\alpha=(-{\partial \omega^k \over \partial E} \big|_{E_k})^{-1/2}$ is the normalization constant. This is actually a general formula for the bound state wave function for the Dirac delta potentials supported by points in $\mathbb{R}^{n}$. For $n=2$, we have   
\be
\psi_k(\mathbf{x}) =  {\alpha \over 2\pi} \; \sum_{i=1}^N  K_0(\sqrt{-E_k}|\mathbf{x}-\mathbf{a_i}|) \, A^{k i}(E_k)  \;.
\label{wavefunction R^2} \ee
Let us recall that the eigenstates for the unperturbed levels are given by unit vectors (\ref{uniteigenvector}),  we write this into the formula for the wave function (\ref{wavefunction R^2}).
As a result, using the first order correction (\ref{1storderA}) to the eigenstate $A^k$ we find that the change of the  original wave function to first order becomes,
\begin{eqnarray}
    \delta \psi_k(x)&=& {(4 \pi E_{B}^k)^{1/2} \over 2\pi}  \sum_{l\neq k } {1\over \ln(|E^B_k|/|E_l^B|)} K_0(\sqrt{|E_B|}|\mathbf{a_k}-\mathbf{a_l}|) K_0(\sqrt{|E^B_k|} |\mathbf{x}-\mathbf{a_l}|) \nonumber\\
    &\sim & \sqrt{2}  |E_B^k|^{1/4} \sum_{l\neq k }{1\over \ln(|E^B_k|/|E_B^l|)} {e^{-\sqrt{|E_B^k|}|\mathbf{a_k}-\mathbf{a_l}|}\over \sqrt{|\mathbf{a_k}-\mathbf{a_l}|}} K_0(\sqrt{|E_B^k|} |\mathbf{x}-\mathbf{a_l}|) \;,
\end{eqnarray}
where we use 
\begin{equation}
    {1\over \Big(-{\partial \omega^k_0(E)\over \partial E}\Big|_{E_B^k}\Big) }=4\pi |E_B^k| \;.
\end{equation}
This form of the wave function clearly shows the tunneling nature of the solutions.
It is now quite straightforward to compute the wave functions in this approximation for all the other cases we consider.

\section*{Conclusion}
\label{Conclusion}

In this paper, we have first reviewed the basic results about some singular interactions, such as the Dirac delta potentials supported by points on flat spaces and hyperbolic manifolds, and delta potentials supported by curves in flat spaces. Moreover, the results in the relativistic extensions of the above-mentioned potentials have been also reviewed which was essentially given in \cite{AltunkaynakErmanTurgut, ErmanTurgut, ErmannumberH, Caglar1, ermangadellauncu, BurakTeo}. The main result of this paper is to develop some kind of perturbation theory applied to a class of singular potentials in order to find the splitting in the energy due to the tunneling. This was only developed extensively  for Dirac delta potentials supported by points in \cite{ErmanTurgut}, here we extend the method to various kinds of Dirac delta potentials as well as to their relativistic versions.

It is possible to give some bounds over the error terms if we assume that the errors in  perturbation theory can be estimated. Typical perturbative expansions are asymptotic therefore a truncation is needed to get more accurate results, one knows that it gets worse beyond a few terms. The more accurate thing to do is to obtain a Borel summed version but that is beyond the content of the present paper, it will depend very much of the specifics of the model whereas we prefer to give a broader perspective.

The comparison with conventional methods certainly would be very useful, nevertheless at present we do not know how a more  conventional approach, such as WKB or instanton calculus can be performed in these singular problems. Since the potentials are localized at points or along the curves, the variation of the potential relative to any wavelength is always much more important. Indeed this unusual behavior changes the problem completely. We need to give a meaning to these potentials first and redevelop the WKB analysis. Our main point here is that in this description of the singular potentials via resolvents, the WKB's reincarnation is given by  a perturbative analysis  of the eigenvalues of the principal operator for large separations of the supports.



\section*{Acknowledgments} 

The authors would like to thank the editors for their invitation to contribute to the special issue and two anonymous referee for his/her valuable comments and suggestions which improve
the paper. First author also acknowledges H. Uncu for his useful suggestion about the numerical calculations in Mathematica.

\end{document}